\documentclass[10pt,twocolumn,superscriptaddress,english,10pt,prl,showpacs,floatfix,aps]{revtex4-1}
\usepackage[latin9]{inputenc}
\usepackage{amsthm}
\usepackage{amsmath}
\usepackage{amssymb}
\usepackage{esint}
\usepackage{graphicx}
\usepackage{upgreek}

\makeatletter

\@ifundefined{textcolor}{}
{%
 \definecolor{BLACK}{gray}{0}
 \definecolor{WHITE}{gray}{1}
 \definecolor{RED}{rgb}{1,0,0}
 \definecolor{GREEN}{rgb}{0,1,0}
 \definecolor{BLUE}{rgb}{0,0,1}
 \definecolor{CYAN}{cmyk}{1,0,0,0}
 \definecolor{MAGENTA}{cmyk}{0,1,0,0}
 \definecolor{YELLOW}{cmyk}{0,0,1,0}
}

\usepackage{soul}
\usepackage{braket}
\usepackage[backref=true,
bookmarksnumbered=true,
bookmarks=true,
bookmarksopen=true,
colorlinks=true,
citecolor=blue,
linkcolor=blue,
anchorcolor=green,
urlcolor=blue,unicode=false]{hyperref}

\let\baraccent=\= 
\renewcommand{\=}[1]{\stackrel{#1}{=}} 

\DeclareMathOperator{\Tr}{Tr}

\makeatother

\usepackage{babel}

\newcommand{\abs}[1]{\left| #1 \right|} 

\begin{document}

\title{Tunneling processes into localized subgap states in superconductors}

\author{Michael Ruby}
\affiliation{\mbox{Fachbereich Physik, Freie Universit\"at Berlin, 14195 Berlin, Germany}}

\author{Falko Pientka}
\affiliation{\mbox{Dahlem Center for Complex Quantum Systems and Fachbereich Physik, Freie Universit\"at Berlin, 14195 Berlin, Germany}}

\author{Yang Peng}
\affiliation{\mbox{Dahlem Center for Complex Quantum Systems and Fachbereich Physik, Freie Universit\"at Berlin, 14195 Berlin, Germany}}

\author{Felix von Oppen}
\affiliation{\mbox{Dahlem Center for Complex Quantum Systems and Fachbereich Physik, Freie Universit\"at Berlin, 14195 Berlin, Germany}}

\author{Benjamin W. Heinrich}
\affiliation{\mbox{Fachbereich Physik, Freie Universit\"at Berlin, 14195 Berlin, Germany}}

\author{Katharina J. Franke}
\affiliation{\mbox{Fachbereich Physik, Freie Universit\"at Berlin, 14195 Berlin, Germany}}

\begin{abstract}
We combine scanning-tunneling-spectroscopy experiments probing magnetic impurities on a superconducting surface with a theoretical analysis of the tunneling processes between (superconducting) tip and substrate. We show that the current is carried by single-electron tunneling at large tip-substrate distances and Andreev reflections at smaller distances. The single-electron current requires relaxation processes between the impurity-induced Shiba bound state and the quasiparticle continuum, allowing us to extract information on such relaxation processes from our analysis.  
\end{abstract}


\maketitle

{\em Introduction.---}Impurity-induced subgap states provide a fruitful window into conventional and unconventional superconductors \cite{rmp,yazdani1,yazdani2}. The Yu-Shiba-Rusinov states \cite{yu,shiba,rusinov} bound by magnetic impurities in conventional $s$-wave superconductors are a simple model system for nonmagnetic impurity resonances in unconventional superconductors, probe the competition between superconducting and Kondo correlations \cite{franke11,yao,bauer13}, and might provide a platform for engineering topological superconducting phases with Majorana end states \cite{bernevig,pientka13,yazdani}. 

In scanning tunneling spectroscopy, Shiba states induce resonances which occur symmetrically at positive and negative bias voltages \cite{yazdani1,yazdani2,franke11, ji08}. Given their subgap nature, it is natural to describe the current into Shiba states as carried by Andreev processes. These processes transfer a Cooper pair into the condensate and are resonantly enhanced by the Shiba state \cite{levy-yeyati,anderson,feigelman,martin13}. Nevertheless, STM experiments on Shiba states are typically analyzed in terms of the tunneling density of states which is appropriate for single-electron tunneling \cite{rmp,flatte,salkola}. This allows one to understand the observed asymmetry in height between the positive- and negative-bias peaks while Andreev processes would necessarily be symmetric in bias (for normal-state tips) \cite{martin13}. 

Here, we combine scanning tunneling microscopy/spectroscopy (STM/STS) of Shiba states using superconducting tips with a comprehensive theoretical analysis to elucidate the nature of the tunneling processes. We show that both single-electron and Andreev tunneling contribute in experiment and explain the observed inversion of peak-height asymmetry as function of tunneling rates. Our analysis shows that STM experiments on Shiba states provide access to quasiparticle relaxation rates in superconductors, complementing recent work on superconducting quantum dots \cite{buitelaar,francesci,marcus} and Josephson junctions \cite{pothier,kos,levy,kozorezov}. 

{\em Experiment.---}We have performed STM experiments probing Mn adatoms on a Pb(111) single crystal surface. The experiments were carried out in a \textsc{Specs} JT-STM at the base temperature of $1.2\,\mathrm{K}$ as well as at $4.8\,\mathrm{K}$. The Pb single crystal surface was cleaned by repeated sputter/anneal cycles until a clean, atomically flat, and superconducting surface was obtained (critical temperature $T_c=7.2\,\mathrm{K}$ and gap $\Delta=1.35\,\mathrm{meV}$ at $1.2\,\mathrm{K}$). Mn adatoms were evaporated onto the clean sample at a temperature below $10$\,K, resulting in a density of $30$ atoms per $100\times 100\,\mathrm{nm}^2$ (see Supplementary Material \cite{supplementary}). Our STM experiments were carried out with a Pb-covered, superconducting tip (see Ref.\ \cite{franke11} for the preparation procedure) which improves resolution far beyond the Fermi-Dirac limit~\cite{bheinrich13,ruby14}. 
 
\begin{figure}[t]
\includegraphics[width=.48\textwidth]{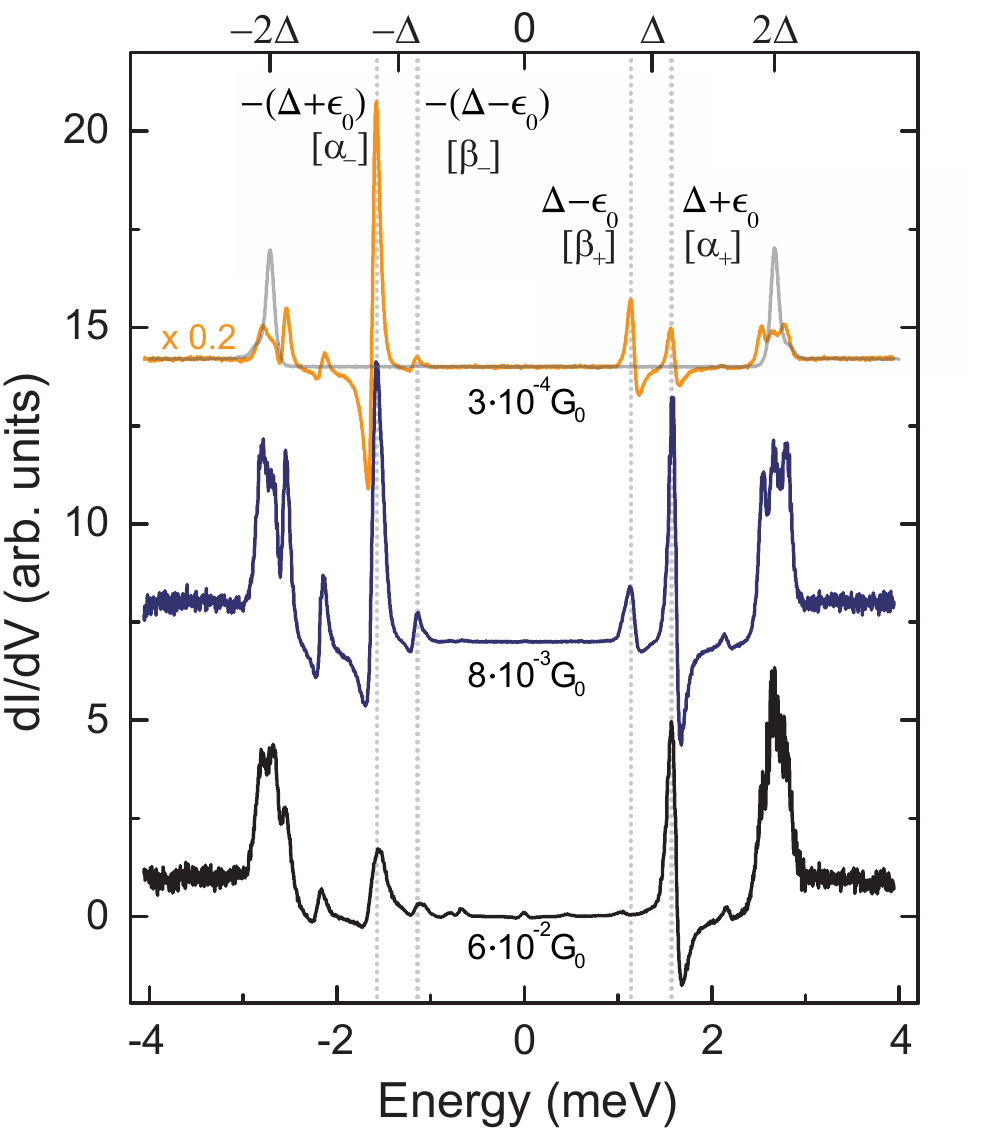}
\caption{d$I/$d$V$ spectra measured on an isolated Mn adatom on Pb(111) for increasing tunneling strength from top to bottom (recorded with a lock-in modulation amplitude of $15\,\mathrm{\upmu V}_\mathrm{rms}$ at a frequency of $912\,$Hz). Spectra are normalized to the `normal-state' conductance measured at $4\,\mathrm{meV}$ (i.e., well outside the superconducting gap) in units of $G_0=2e^2/h$, offset for clarity, and scaled when indicated for better visibility. The distance to the  closest neighboring Mn atom was greater than $5$\,nm. A spectrum acquired above the clean Pb(111) surface overlays the smallest-conductance trace (top curve) for comparison. The four peaks originating from the deepest Shiba level are marked by dashed lines at $e|V|=\pm (\Delta\pm\epsilon_0)$.}
\label{fig:didv}
\end{figure}

Figure~\ref{fig:didv} shows spectra of the differential conductance d$I/$d$V$ as a function of bias $V$, acquired at various tip--sample distances and thus tunneling strengths with the tip placed above a Mn adatom. All spectra share the same characteristic peaks \cite{note1} but their intensities (normalized to the normal-state conductance) depend strongly on the tunneling strength and the sign of the bias voltage. 

\begin{figure}[t]
\includegraphics[width=.15\textwidth]{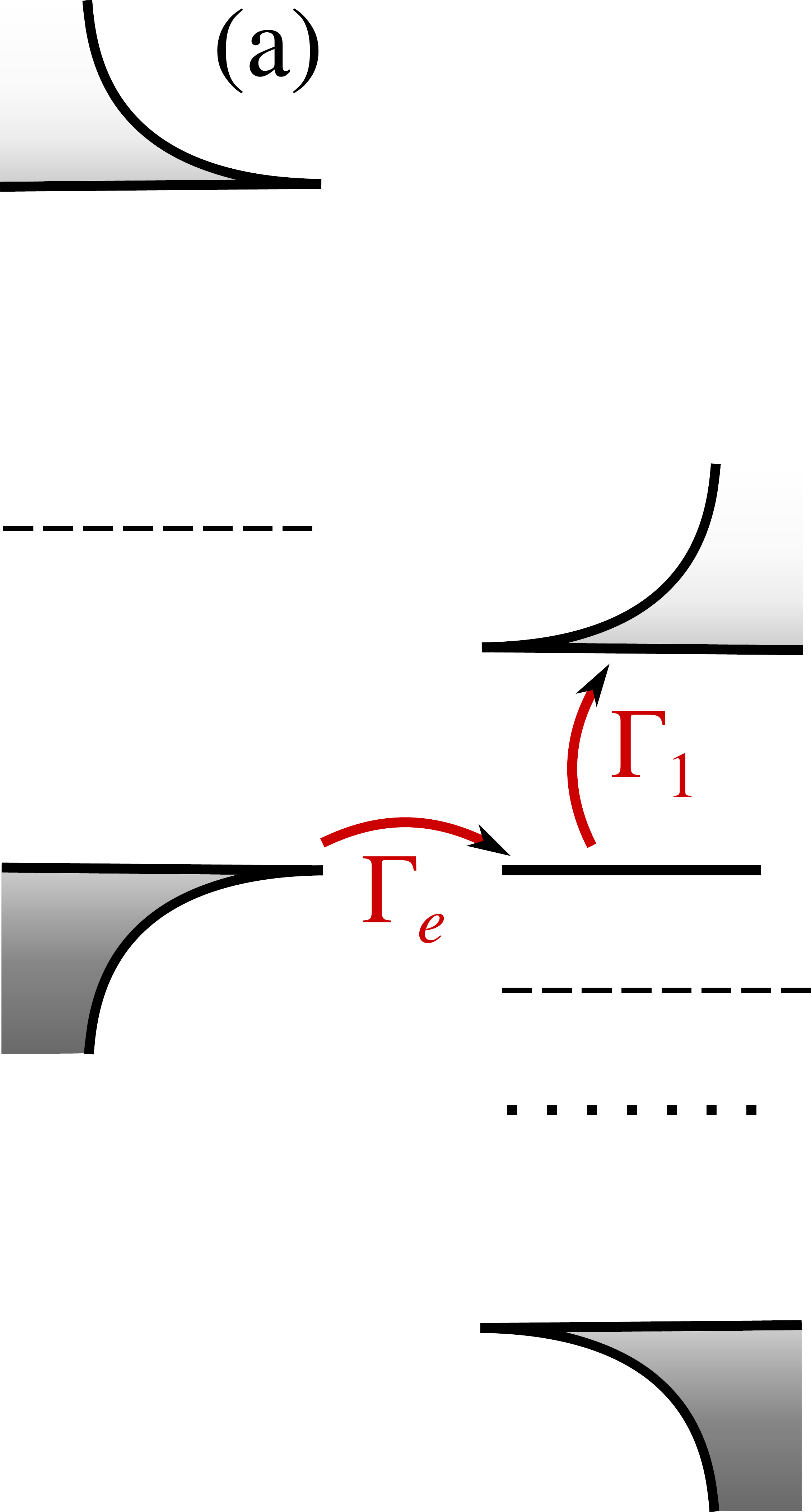}
\hspace{.2em}
\includegraphics[width=.15\textwidth]{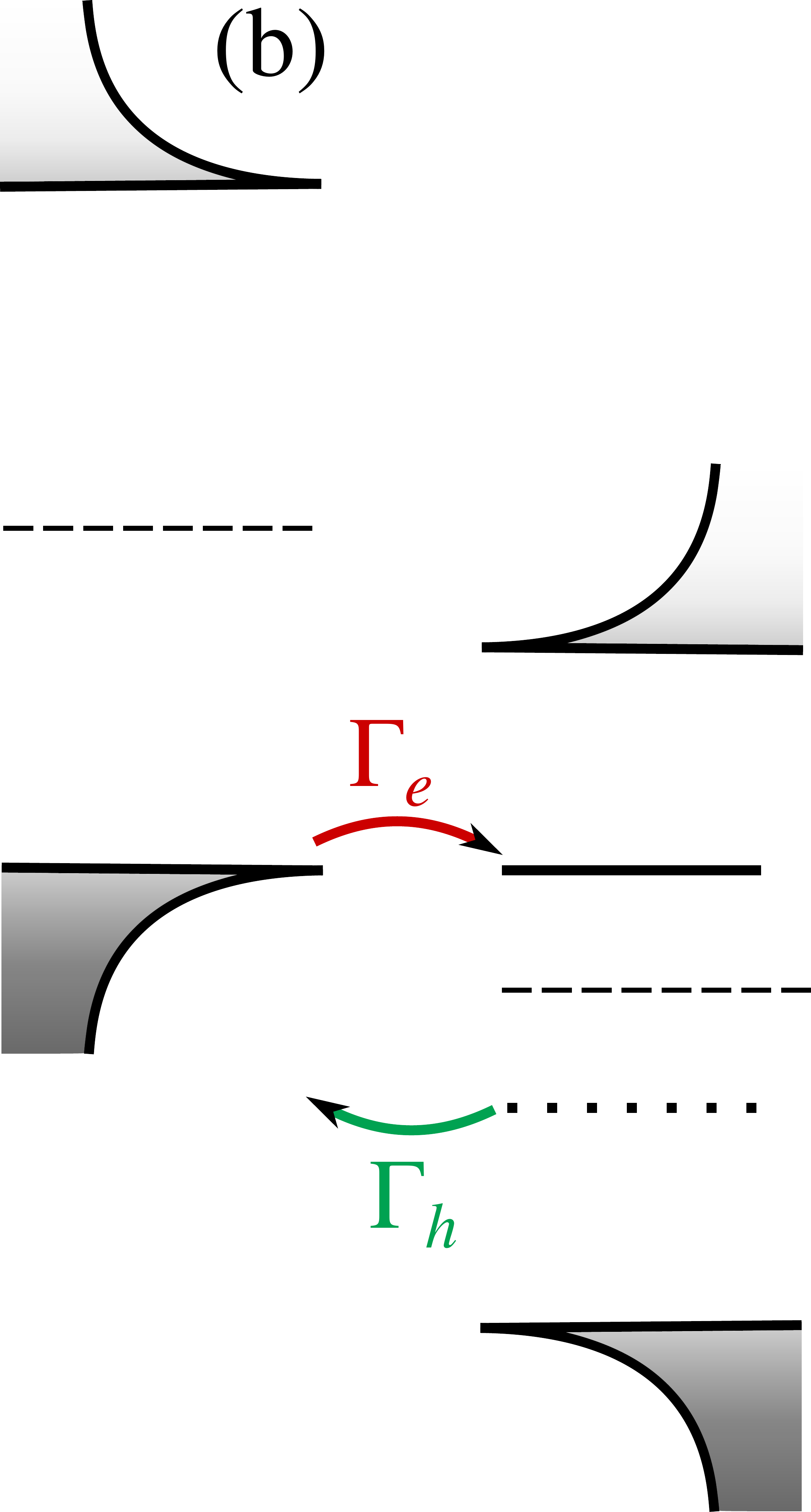}
\hspace{.2em}
\includegraphics[width=.15\textwidth]{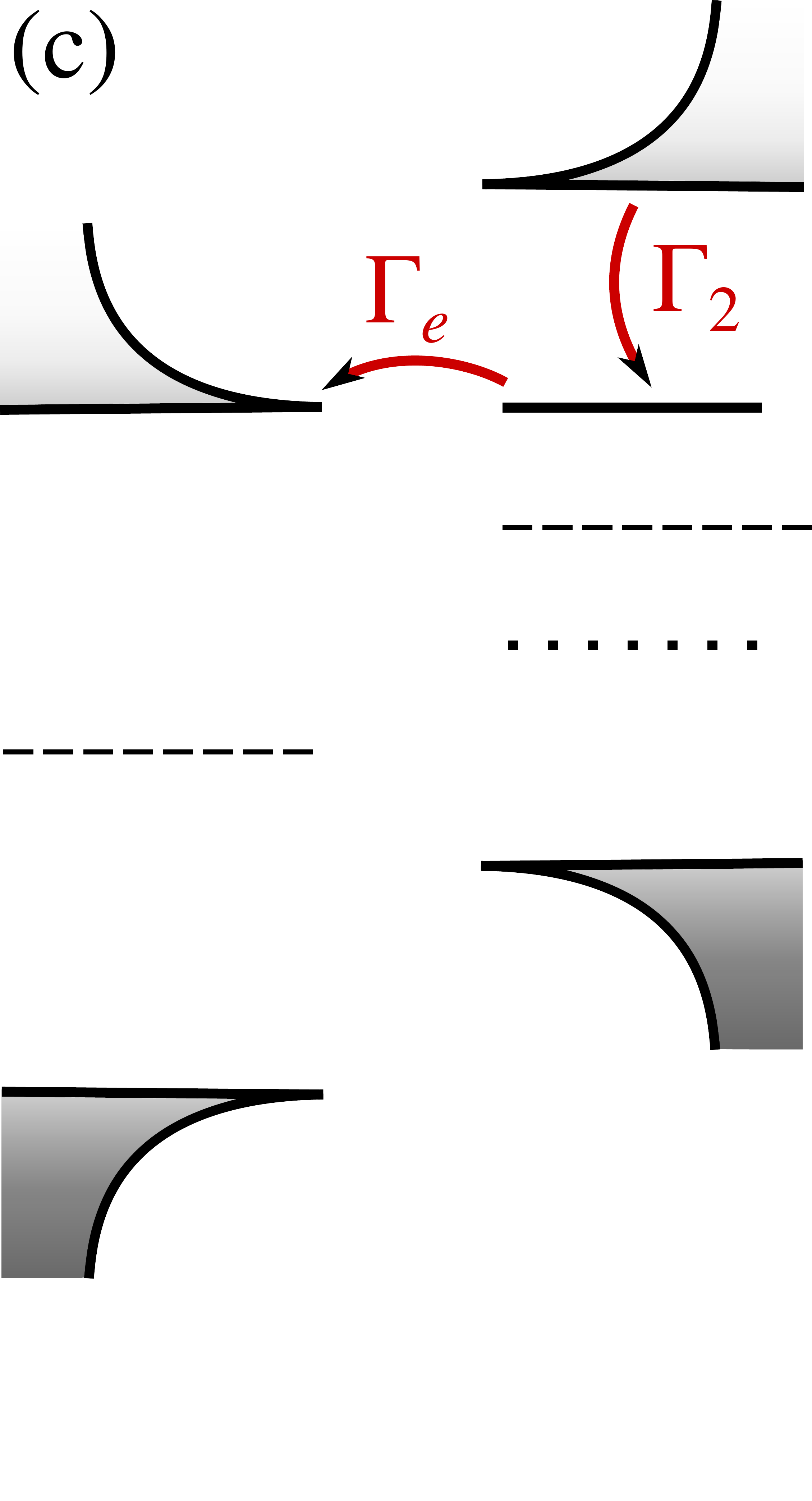}
\caption{Principal tunneling processes involving a Shiba state (solid line) within the superconducting gap (enclosed by BCS quasiparticle peaks). The chemical potential is represented by a dashed line.  (a) Single-electron tunneling from tip to substrate (rate $\Gamma_e(\omega)$) with subsequent relaxation from the Shiba state to the quasiparticle continuum (rate $\Gamma_1$). (b) Andreev process transfering a Cooper pair to the substrate by electron and hole tunneling (with rates $\Gamma_e(\omega)$ and $\Gamma_h(\omega)$, respectively). The processes in (a) and (b) both contribute near the threshold $eV=\Delta+\epsilon_0$.  (c) Single-electron tunneling from substrate to tip (with rate $\Gamma_e(\omega)$) after occupation of the Shiba state by the relaxation of a thermal quasiparticle (with rate $\Gamma_2$), contributing to the thermal peak at $eV=-(\Delta-\epsilon_0)$. The current at the other two thresholds $eV=-(\Delta+\epsilon_0)$ and $eV=\Delta-\epsilon_0$ is carried by analogous hole processes (see \cite{supplementary}).}
\label{fig:tunnel_process}
\end{figure}

The peaks in the d$I/$d$V$ spectra appear at thresholds for various fundamental tunneling processes between superconducting tip and substrate with magnetic adatom: (i) Single electrons can tunnel when the negative-energy quasiparticle continuum of the tip overlaps with the positive-energy continuum of the substrate (or vice versa). This requires a threshold voltage $eV=\pm 2\Delta$. (ii) Thermally excited quasiparticles (holes) in the positive-(negative-) energy quasiparticle continuum induce a single-particle current even near zero bias. (iii) With a Shiba state of energy $\epsilon_0$, a single-particle current flows when the negative-energy continuum of the tip overlaps with the Shiba state, or the positive-energy continuum with the symmetric energy $-\epsilon_0$. These processes have threshold biases $eV=\pm(\Delta+\epsilon_0)$. (iv) Due to thermal occupation, a single-electron current can also flow when the positive-energy continuum overlaps with the Shiba state (and symmetrically when the negative-energy continuum overlaps with $-\epsilon_0$). This requires a threshold bias $eV=\pm(\Delta-\epsilon_0)$. (v) At $e|V|<2\Delta$, an electron from, say, the tip can be reflected as a hole, transferring a Cooper pair. As all tunneling electrons and holes gain an energy $eV$, (multiple) Andreev processes between the quasiparticle continua have thresholds $eV=\pm 2\Delta/n$ with $n=2,3,\ldots$. Andreev processes require two or more particles to cross the tunnel barrier and thus become relevant for strong tunnel coupling only \cite{ternes06}. (vi) Shiba states induce additional {\em resonant} Andreev processes which become relevant at much lower tunneling rates. An electron from the negative-energy continuum of the tip can virtually tunnel into the Shiba state, reflect as a hole, and resonantly transfer a Cooper pair into the condensate of the substrate. Together with a similar process at reverse bias, this leads to thresholds at $eV=\pm(\Delta+\epsilon_0)$ which {\em coincide} with those for the single-electron processes. The principal tunneling processes involving the Shiba states are sketched in Fig.\ \ref{fig:tunnel_process}.

There is an important difference between the single-electron and resonant Andreev processes \cite{martin13}. Single-electron processes change the occupation of the Shiba state while Andreev processes merely transfer Cooper pairs into the condensate. Thus, a continuous current flow by single-electron processes requires relaxation processes which empty the Shiba state after it is occupied from the tip (or occupy the empty Shiba state), see Fig.~\ref{fig:tunnel_process}. At finite temperature, a quasiparticle in the Shiba state can be excited to the continuum by absorption of a phonon or a photon (with rate $\Gamma_1$). Conversely, a thermally excited quasiparticle can relax into the Shiba state by emission (with rate $\Gamma_2$). 

The observed peaks in the d$I/$d$V$ spectra can now be correlated with Shiba states of energy $\simeq$\,0.22\,meV, $\simeq$\,0.77\,meV, and $\simeq$\,1.18\,meV, respectively. The multiple Shiba states may reflect different angular-momentum channels or spin states $S>1/2$~\cite{ji08, moca08,zitko11}. To analyze the tunneling processes, we focus on the most intense Shiba state at $\epsilon_0\simeq 0.22\,\mathrm{meV}$. This state not only leads to the two {\em main} peaks at $eV=\pm(\Delta +\epsilon_0)$ (with peak height $\alpha_\pm$), but  also to two pronounced \textit{thermal} peaks at $eV=\pm(\Delta -\epsilon_0)$ (with peak height $\beta_\pm$). As it is the deepest state, its theoretical interpretation turns out to be least affected by the presence of the other Shiba states.

\begin{figure}[t]
\includegraphics[width=.48\textwidth]{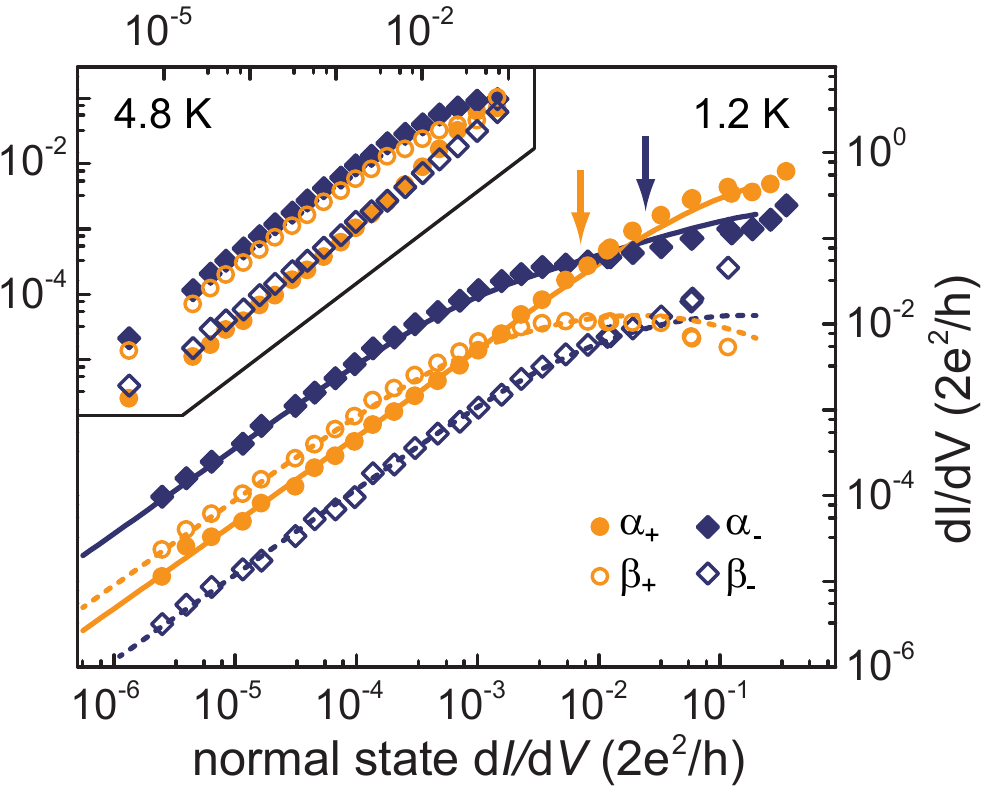}
\caption{Peak heights $\alpha_\pm$ and $\beta_\pm$ of the four resonances associated with the deepest Shiba level (marked by dashed lines in Fig.~\ref{fig:didv}) as a function of normal state conductance at $T=1.2\, {\rm K}$ (main panel) and $T=4.8\, {\rm K}$ (inset). The full (dashed) lines are fits to Eqs.\ (\ref{IR}) and (\ref{IA}) for the main (thermal) peaks. 
The crossover points between single-electron and Andreev contributions to $\alpha_\pm$ are indicated by arrows.}
\label{fig:peak_heights}
\end{figure}

The heights of the peaks associated with this Shiba state are plotted in Fig.~\ref{fig:peak_heights} over several decades in normal-state tunneling conductance. We draw attention to two important features of these data. First, the peak heights vary linearly over a wide region before turning sublinear at larger tunneling rates. Second, the asymmetry in the peak heights $\alpha_\pm$ between positive and negative biases inverts as a function of tunneling strength: At small tunneling rates, $\alpha_+<\alpha_-$, while at large tunneling rates, $\alpha_+>\alpha_-$. It is also evident that the inversion of the peak heights occurs at the crossover between the linear and sublinear regimes.

{\em Theoretical analysis.---}It is often assumed \cite{rmp} that the peak heights at positive and negative biases measure the electron and hole components $u$ and $v$ of the Shiba wavefunction. The observed inversion of peak heights implies that this cannot hold in general. To gain further insight, we calculate the subgap current theoretically by a standard Keldysh calculation \cite{Haug2008,cuevas96} (see Supplementary Material for details \cite{supplementary}). Here, we focus on the physics underlying the results. Our calculation includes single-electron and Andreev processes involving the Shiba state as well as phenomenological rates $\Gamma_1$ and $\Gamma_2$ for relaxation processes between Shiba state and quasiparticle continuum. We neglect the non-resonant Andreev reflections at the superconducting tip (and thus multiple Andreev reflections \cite{averin}), which is justified except in the regime of very strong tunneling. With this approximation, the tunneling current becomes a sum of single-particle and Andreev currents, $I=I^s + I^a$, with 
\begin{align}
 I^s=&\ {e}\int \frac{d\omega}{2\pi\hbar} \, \biggl\{\frac{\Gamma_1 [\Gamma_e n_F(\omega_-)-\Gamma_hn_F(\omega_+)]}{(\omega-\epsilon_0)^2+(\Gamma/2)^2}\nonumber\\
 -&\ \frac{\Gamma_2 [\Gamma_e (1-n_F(\omega_-))-\Gamma_h(1-n_F(\omega_+))]}{(\omega-\epsilon_0)^2+(\Gamma/2)^2}\biggr\},
 \label{IR}\\
 I^a=&\ {2e}\int \frac{d\omega}{2\pi\hbar} \frac{\Gamma_h \Gamma_e[n_F(\omega_-)-n_F(\omega_+)]}{(\omega-\epsilon_0)^2+(\Gamma/2)^2}.
 \label{IA}
\end{align}
Here, the Fermi functions $n_F$ are evaluated at $\omega_\pm=\omega\pm eV$ and $\Gamma=\Gamma_e+\Gamma_h+\Gamma_1+\Gamma_2$. 

The expressions for $I^s$ and $I^a$ can be understood in terms of the basic processes discussed above. The Andreev current $I^a$ involves tunneling of an electron, described by $\Gamma_e(\omega)=2\pi u^2 \rho(\omega\,-\,eV)t^2$, and a hole, described by $\Gamma_h(\omega)=2\pi v^2 \rho(\omega\,+\,eV)t^2$.  Here, $t$ is the amplitude for tunneling between tip and substrate. The rates $\Gamma_e$ and $\Gamma_h$ are strongly $\omega$-dependent through the tip's BCS density of states $\rho(\omega)$. The denominator in Eq.\ (\ref{IA}) reflects the intermediate virtual occupation of the Shiba state. It includes the rates $\Gamma_1$ for depopulating the Shiba state by excitation to the continuum and $\Gamma_2$ for occupying the Shiba state by a thermally excited quasiparticle. The latter processes are assumed $\omega$-independent. The four contributions to the single-particle current $I^s$ directly correspond to the peaks $\alpha_+$ [term $\propto \Gamma_1\Gamma_e$, see Fig.\ \ref{fig:tunnel_process}(a)], $\alpha_-$ (term $\propto \Gamma_1\Gamma_h$), $\beta_-$ [term $\propto \Gamma_2\Gamma_e$, see Fig.\ \ref{fig:tunnel_process}(c)], and $\beta_+$ (term $\propto \Gamma_2\Gamma_h$). 

Eqs.\ (\ref{IR}) and (\ref{IA}) provide the following basic picture consistent with the data in Fig.\ \ref{fig:peak_heights}: At weak tunneling, the relaxation rates $\Gamma_1$ and $\Gamma_2$ are faster than the tip-substrate tunneling. Once an electron tunnels into the Shiba state from the tip, it is rapidly excited to the quasiparticle continuum. In this regime, the tunnel current is dominated by the single-electron current $I^s$ which is proportional to $t^2$ and thus to the normal-state conductance. The Andreev current $I^a$ is a small correction scaling as $t^4$. This explains the wide linear regime in Fig.\ \ref{fig:peak_heights}. At stronger tunneling, the tunneling rates become comparable to and eventually larger than the relaxation rates $\Gamma_1$ and $\Gamma_2$. Here, the $t$-dependence of the broadening $\Gamma$ leads to a sublinear or even a decreasing dependence of the peak heights on the normal-state conductance. 
As the relaxation processes are thermally activated, the crossover point between linear and sublinear regime is strongly temperature dependent, moving to lower normal-state conductances for lower temperatures. This is consistent with a comparison between main panel and inset of Fig.~\ref{fig:peak_heights}. 

{\em Linear regime.---}This picture is substantiated by a quantitative analysis of the linear regime. For weak tunneling and $\Gamma_1\gg\Gamma_2$ (i.e., $\epsilon_0\gg T$), Eq.\ (\ref{IR}) yields \cite{supplementary}
\begin{align}
  \alpha_+ \sim \frac{2e^2}{h}\frac{\gamma_e \sqrt{\Delta}}{(\Gamma_1)^{3/2}} \,\,\,\, ; \,\,\,\, 
\beta_- = \alpha_+\frac{\Gamma_2}{\Gamma_1} \label{alpha_peak}
\end{align}
for the peak heights. Here, we introduced the normal-state electron (hole) tunneling rate $\gamma_e = 2\pi t^2 \nu_0 u^2$ ($\gamma_h = 2\pi t^2 \nu_0 v^2$), where $\nu_0$ is the normal-state density of states of the tip. The expressions for $\alpha_-$ and $\beta_+$ simply differ by the substitution $u\leftrightarrow v$ (or $\gamma_e \leftrightarrow \gamma_h$). Thus, in this regime, the peak height is indeed a measure of the Shiba wavefunction at the tip position. From the data in Fig.~\ref{fig:peak_heights}, we extract $\alpha_+/\alpha_-=(u/v)^2 \approx 0.13$. 

All four peaks are related by the relation $\alpha_+\beta_+=\alpha_-\beta_-$. This is readily checked against the data in Fig.~\ref{fig:peak_heights} and indeed, we find that this identity is well satisfied in the linear regime \cite{supplementary}. Moreover, the thermal and main peaks in Eq.~(\ref{alpha_peak}) differ only by a ratio of relaxation rates, ${\alpha_+}/{\beta_-} ={\Gamma_1}/{\Gamma_2} =\exp(\epsilon_0/T)$. Here, the last equality follows from detailed balance. This is in excellent agreement for the data at $T=4.8\, {\rm K}$. At $T=1.2\, {\rm K}$, we extract a slightly higher temperature of $T=1.6\, {\rm K}$ from the ratio of peak heights. Still, these considerations point to a relaxation process involving thermal activation rather than the quasiparticle bath suggested in Ref.\ \cite{martin13}. 

As $\Gamma_1$ increases with temperature, Eq.~(\ref{alpha_peak}) also predicts the peak heights in the linear regime to decrease with $T$. This is consistent with the data as seen by comparing the main panel and the inset in Fig.~\ref{fig:peak_heights}.

{\em Regime of strong tunneling.---}As tunneling rates increase relative to the relaxation rates, the magnitude of the single-particle conductance reaches a maximum and eventually decreases. As a result, the thermal peaks $\beta_\pm$ should exhibit a maximum vs normal-state conductance. The situation is different for the main peaks $\alpha_\pm$ with their additional Andreev contribution which keeps increasing and eventually dominates the peak magnitude. Sufficiently far into this regime, Eqs.~(\ref{IR}) and (\ref{IA}) yield
\begin{align}
  \alpha_+ &\sim ({2e^2}/{h})(\gamma_h\sqrt{\Delta/\epsilon_0}/(\gamma_e\sqrt{\Delta})^{2/3}) 
\label{alpha_peak1}\,, \\ 
  \beta_- &\sim ({2e^2}/{h})({\Gamma_2}/{(\gamma_e\sqrt{\Delta})^{2/3}}) 
  \label{alpha_peak2}
\end{align}
as well as $\alpha_-$ and $\beta_+$ which differ again by $u\leftrightarrow v$. The main peaks $\alpha_\pm$ keep increasing with tunneling $\gamma_{e,h}$, albeit with a sublinear dependence.  We note that the transition from weak to strong tunneling is also accompanied by a change in the lineshape of the peaks (see \cite{supplementary}). 

Unlike for normal-metal tips \cite{martin13,supplementary}, the Andreev contribution to the main peaks $\alpha_\pm$ is asymmetric for a superconducting tip, but with the asymmetry reversed relative to single-electron tunneling. While we have $\alpha_+/\alpha_-=(u/v)^2$ in the linear regime, Eq.\ (\ref{alpha_peak1}) predicts $\alpha_+/\alpha_-= (v/u)^{10/3}$ in the Andreev-dominated regime. Indeed, an inversion of the peak heights $\alpha_\pm$ is seen in Fig.\ \ref{fig:peak_heights}, as pointed out above. 

Eq.\ (\ref{alpha_peak2}) predicts that also the thermal peaks invert, from $\beta_-/\beta_+=(u/v)^2$ in the linear regime to $\beta_-/\beta_+=(v/u)^{4/3}$ in the sublinear regime. This inversion is consistent with the data  in Fig.\ \ref{fig:peak_heights}. In adddtion, theory predicts the thermal peaks to assume a maximum as a function of normal state conductance. We observe such a maximum only for $\beta_+$. For $\beta_-$, the peak is expected to occur only at rather large normal-state conductance where our approximations of neglecting multiple Andreev reflections and a peak width smaller than $\epsilon_0$ break down. 

To further substantiate our analysis, we have used Eqs.\ (\ref{IR}) and (\ref{IA}) to fit all four peaks $\alpha_\pm$ and $\beta_\pm$ over the entire range of tunneling strengths, see Fig.\ \ref{fig:peak_heights} \cite{supplementary}. There is excellent agreement between theory and experiment. We attribute the deviations for $\beta_-$ at large normal-state conductance to additional contribution from multiple Andreev reflections. We can also extract the normal-state conductance at which the Andreev and single-particle contributions to the main peaks become comparable, see the arrows in Fig.\ \ref{fig:peak_heights}. (Note that this is distinct from the crossover between linear and sublinear dependence.) For $\alpha_+$, this happens when $2\Gamma_h(2\epsilon_0) \sim \Gamma_1$, and for $\alpha_-$, when $2\Gamma_e(2\epsilon_0)\sim\Gamma_1$. As $v^2>u^2$, the Andreev contribution sets in considerably earlier for $\alpha_+$ than for $\alpha_-$.

{\em Relaxation rates.---}To extract the relaxation rates $\Gamma_1$ and $\Gamma_2$ quantitatively, we focus on the thermal peak $\beta_+$. In the sublinear regime it contributes a current $I \sim e\Gamma_2/\hbar$ \cite{supplementary}. Moreover, Eq.\ (\ref{alpha_peak}) predicts $\Gamma_1 = (\alpha_+/\beta_-)\Gamma_2$ in the linear regime.  We can thus extract both relaxation rates directly from the experimental data. This yields $\Gamma_1$  of order $0.2\,\mathrm{ns}$ at $T=1.2\,\mathrm{K}$ and $6\,\mathrm{ps}$ at $T=4.8\,\mathrm{K}$. At the lower temperature, $\Gamma_2$ differs appreciably from $\Gamma_1$ and we extract a value of order $0.7\,\mathrm{ns}$. While in principle, one could also rely on the main peaks to extract $\Gamma_1$ and $\Gamma_2$, this is less accurate since Andreev current and thermal peaks also contribute to the total current. 

If the relaxation process relied on directly exciting a quasiparticle from the Shiba state to the continuum, we would predict a ratio of the relaxation rates at the two experimental temperatures of order $\sim 10^4$. We can account for a substantial part of the apparent discrepancy with our observations by recalling that there are additional Shiba states. If relaxation proceeds as a multistep process which first excites to the second Shiba state, we predict a ratio of relaxation rates which is consistent with experiment (see also \cite{supplementary}). 

{\em Conclusions.---}We show that STM experiments on subgap states in superconductors probe both single-electron and Andreev tunneling. We emphasize that such experiments are particularly fruitful when performed with superconducting tips. In this case, thermal smearing can be neglected and the temperature dependence of the current arises entirely from activated quasiparticle relaxation processes. Moreover, the additional thermal peaks facilitate the analysis and provide access to the relaxation rates. We find that at weak tip-substrate tunneling, the current is dominated by single-electron tunneling and linear in the normal-state conductance. This regime can be used to map out the bound-state wavefunction. At stronger tip-substrate tunneling, the dependence on the normal-state conductance becomes sublinear. While the dependence on the Shiba wavefunction becomes more involved, this regime provides access to pertinent quasiparticle relaxation rates involving the subgap states. Specifically, we can extract the rates for quasiparticle relaxation into and out of the bound state. The present experiment was restricted to two different temperatures. To gain further microscopic understanding of the relaxation processes, it would be rewarding to perform more systematic experiments as a function of temperature. 

\begin{acknowledgments}
We thank Piet Brouwer and Leonid Glazman for discussions, and acknowledge financial support by the Deutsche Forschungsgemeinschaft through SFB 658 and FR2726/4 (KF) as well as SPP 1285 and SPP 1666 (FvO), by an ERC grant NanoSpin (KF) and the Helmholtz Virtual Institute {\em New States of Matter and Their Excitations} (FvO).
\end{acknowledgments}

\clearpage

\renewcommand{\theequation}{S\arabic{equation}}
\renewcommand{\thefigure}{S\arabic{figure}}

\onecolumngrid

\section*{Supplementary Material\\Tunneling processes into localized subgap states in superconductors}

\section{Experimental Details}

\subsection{\texorpdfstring{Topography of $\protect \mathrm{Mn}$ Adatoms on $\protect \mathrm{Pb}(111)$}{}}
\begin{figure}[bth]
\includegraphics[width=.85\textwidth]{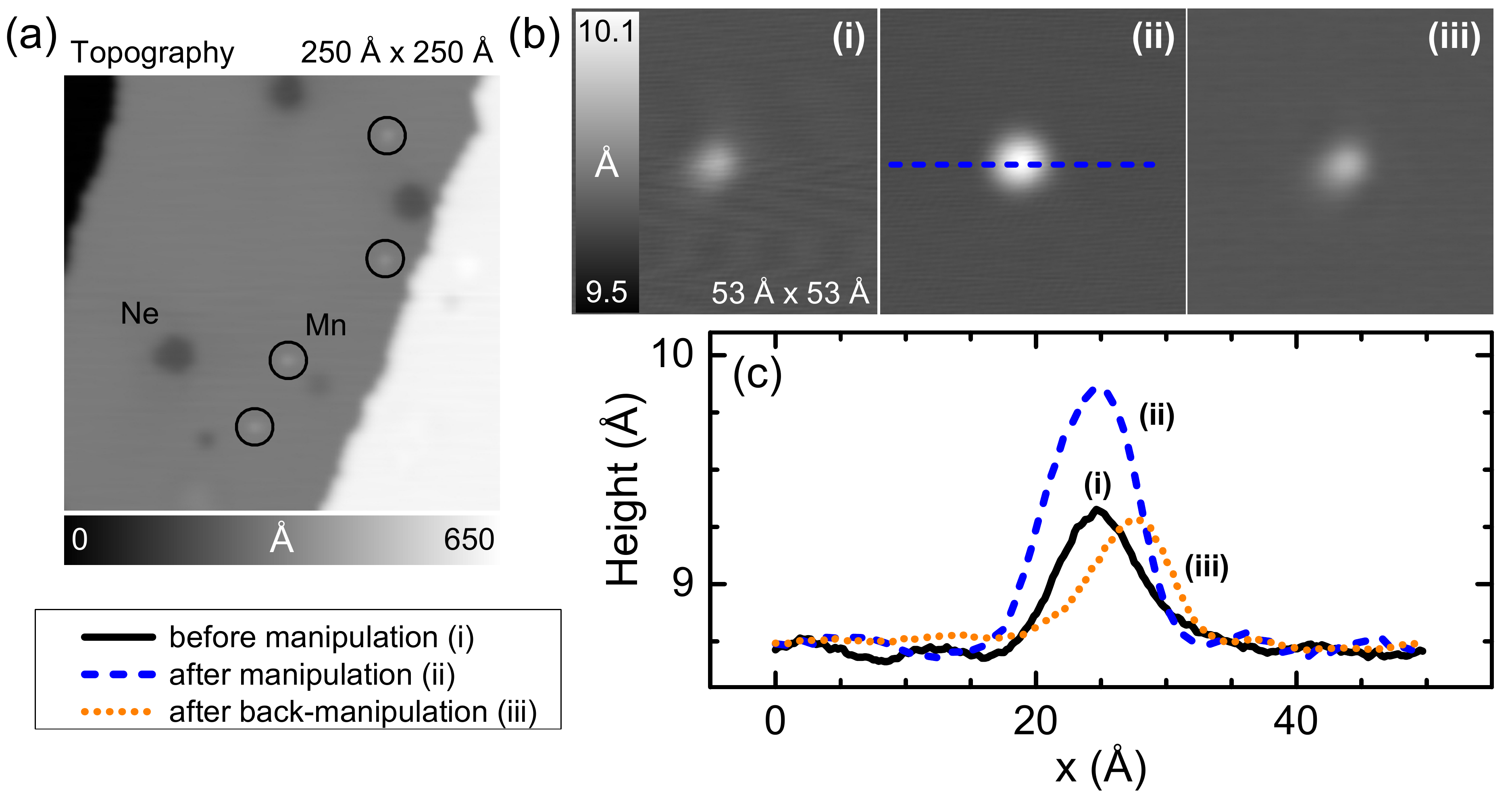}
\caption{
(a) Topography of three terraces of the Pb(111) surface with an evaporated Mn density of $\approx 30$ adatoms per $100\,\times\,100\,\mathrm{nm}^2$. Mn adatoms are marked by black circles. The dark depletions of different sizes are nonmagnetic Neon sub-surface inclusions, that originate from the sample cleaning process. Setpoint: $50\,\mathrm{mV}, 200\,\mathrm{pA}$.
(b) Topography of a single Mn adatom in its two different adsorption states.
(c) Line profiles across the atom for all adsorption states. The lateral shift of the adatom after the back-manipulation (orange, dotted) is due to jumping into a neighboring adsorption site equivalent to the initial one.}

\label{Sfig:Topography}
\end{figure}

The Pb(111) surface was cleaned by $\mathrm{Ne}^+$ sputtering, followed by annealing to $T=430\,\mathrm{K}$.
This yields an atomically clean surface with terraces of several nm width, which are separated by mono-atomic steps [see Fig.~\ref{Sfig:Topography}(a)]. Residual Ne atoms from sputtering form nonmagnetic nano-cavities below the surface, which appear as hexagonal protrusions or depletions of different sizes in topography \cite{ruby14}. They do not show any signatures of subgap resonances.

Manganese atoms were deposited on the clean Pb(111) sample inside the STM at a temperature below $10\,\mathrm{K}$. All Mn adatoms have the same apparent height after the evaporation. By contacting the adatom with the tip at a bias of $5\,\mathrm{mV}$, we can induce a change in the adsorption configuration. The resulting species has a larger apparent height [Fig.~\ref{Sfig:Topography}(b) and (c)]. The manipulation is reversible: contact formation at a bias of $-180\,\mathrm{mV}$ results in the initial apparent height. After back-manipulation the adatom is shifted laterally with respect to the initial position [see line-profiles in Fig.~\ref{Sfig:Topography}(c)]. Thus, the manipulation controllably changes the adsorption site of the adatom. We took care that the absolute tip height did not change during the manipulation to rule out a change of the tip apex. 

Both configurations show distinct d$I$/d$V$ spectra. The initial adsorption site has been investigated by Ji \textit{et al.}~\cite{ji08}, showing multiple Shiba states. Our spectra on this species show the same characteristic features. 
In the main manuscript we focus on the higher species, because it is stable upon tip approach at low bias and thus allows us to investigate d$I$/d$V$ spectra over a large conductance range.

\subsection{\texorpdfstring{$\protect \mathrm{d}I/\mathrm{d}V$ spectra on a $\protect \mathrm{Mn}$ adatom at $4.8\,$K}{}}

\begin{figure}[bth]
\includegraphics[width=.48\textwidth]{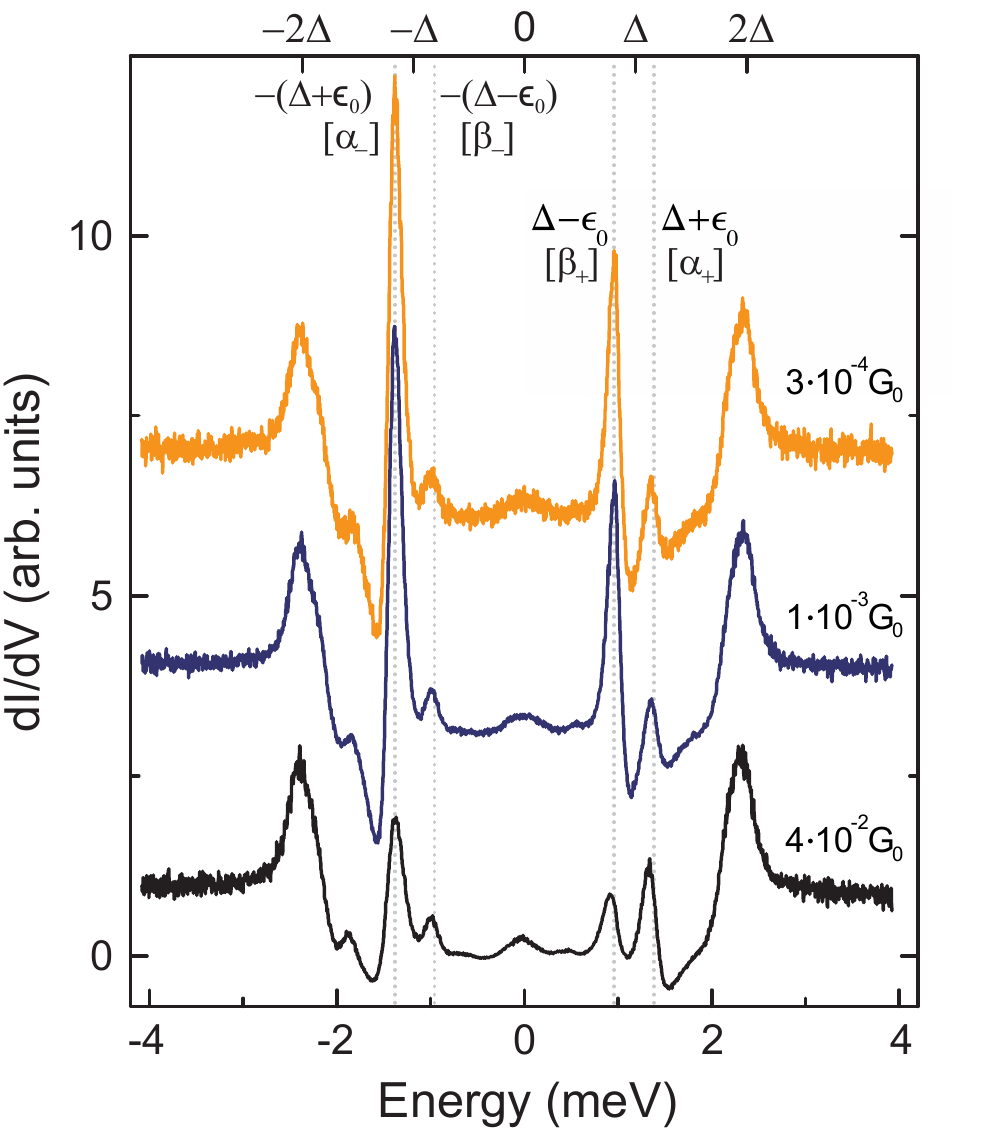}
\caption{d$I/$d$V$ spectra acquired on a Mn adatom at $4.8\,$K ($912\,$Hz, $35\,\mathrm{\upmu V}_\mathrm{rms}$). Spectra are normalized to the normal-state conductance (indicated in the graph). The figure includes assignments of the peaks to the main peaks ($\alpha_\pm$) and the thermal peaks ($\beta_\pm$).}  
\label{Sfig:didv_4.8K}
\end{figure}

In Fig.~\ref{Sfig:didv_4.8K} we show three examples of d$I/$d$V$ spectra at the higher temperature of $4.8\,$K at different tip--sample distances, \textit{i.e.}, different tunneling strengths. At this temperature, only two Shiba states are well resolved due to the increased width of the resonances. An additional zero-energy resonance is observed due to tunneling of thermally excited quasiparticles in tip and sample [process (ii), as described in the main text]. The peak heights $\alpha_+$, $\alpha_-$, $\beta_+$, and $\beta_-$ as shown in the inset of Fig.~3 were extracted from these spectra. Notice that the relative intensities of $\alpha_{\pm}$ change with increasing tunneling strength. However, unlike at 1.2\,K, we do not observe a full inversion of peak heights due to the larger relaxation rate. 

\subsection{\texorpdfstring{Experiments with a normal-metal tip at $1.2\,$K}{}}

\begin{figure}[bth]
\includegraphics[width=.99\textwidth]{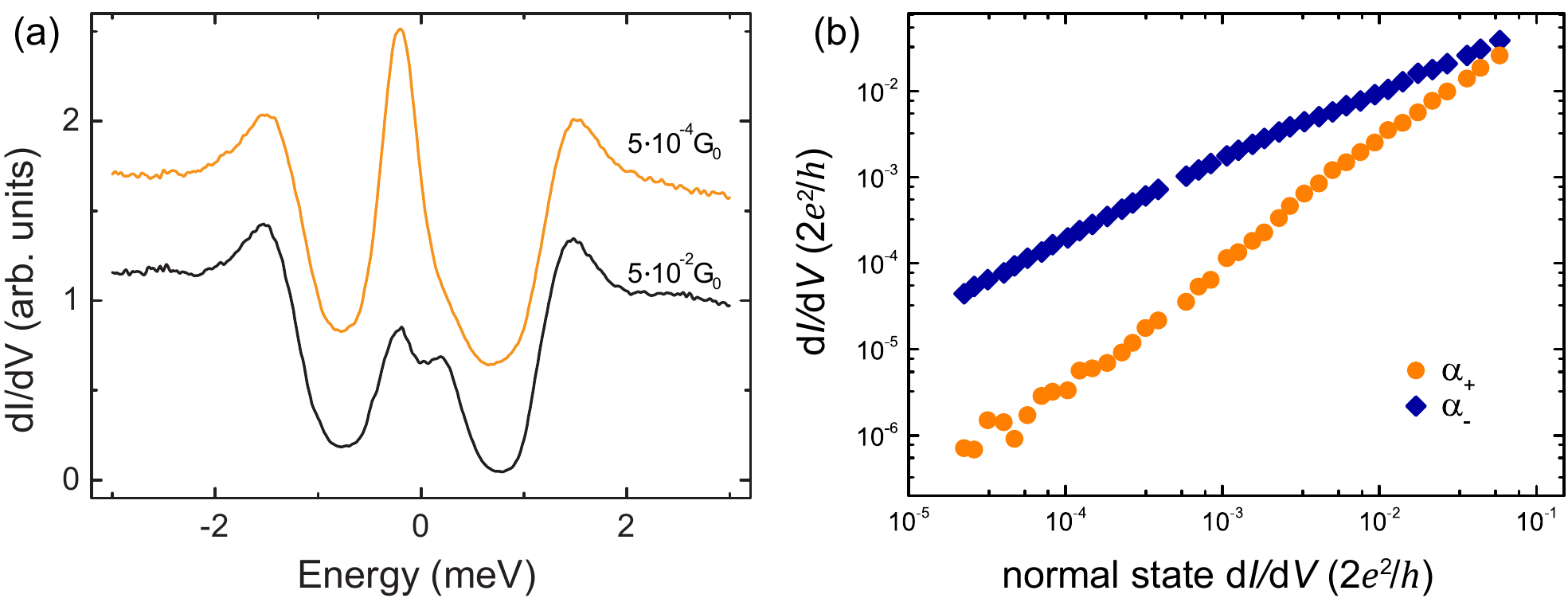}
\caption{(a) Two d$I/$d$V$ spectra acquired on a Mn adatom at $1.2\,$K with a normal-metal tip ($912\,$Hz, $80\,\mathrm{\upmu V}_\mathrm{rms}$). Spectra are normalized to the normal-state conductance, as indicated in the graph.
(b) Peak heights $\alpha_+$ and $\alpha_-$ of the Shiba level as a function of normal state conductance at $T=1.2\, {\rm K}$, measured with a normal-metal tip. 
}
\label{Sfig:didv_metal}
\end{figure}

In the main manuscript, we focus on experiments with a superconducting tip. For completeness, we include spectra and corresponding peak-height \textit{vs.} conductance curves acquired with a normal-metal tip at $1.2\,$K (Fig.~\ref{Sfig:didv_metal}). Due to thermal broadening of the tip's Fermi edge ($\approx\,360\,\mathrm{\upmu V}$), the energy resolution is drastically decreased compared to measurements with superconducting tips and only one pair of Shiba resonances is resolved in Fig.~\ref{Sfig:didv_metal}(a). From a set of such spectra, we extract the peak heights $\alpha_{\pm}$ [Fig.~\ref{Sfig:didv_metal}(b)]. In agreement with Ref.\ \cite{martin13}, we observe an asymmetry  in the weak coupling regime, which reduces when approaching the Andreev regime and reaches (almost) equal intensity for the strongest coupling accessible in the experiment.

%
%

\section{Theoretical details}

Here, we derive the expressions for the tunneling current between a superconducting tip and a superconducting sample with magnetic impurity, as given in Eqs.~(1) and (2) of the main text. We apply the nonequilibrium Green function method used in \cite{cuevas96}.

\subsection{Green-function expression for the current}

The system is described by the Hamiltonian $\hat{H}=\hat{H}_{L}+\hat{H}_{R}+\hat{H}_{T}$, where the three parts describe the tip, the substrate, and the tunnel coupling.
The superconducting tip is described by the BCS Hamiltonian 
\begin{align}
 \hat{H}_{L}=\int \frac{d\mathbf{k}}{(2\pi)^3} \biggl[\sum_\sigma \xi_k   \hat{c}_{L,\mathbf{k}\sigma}^\dag  \hat{c}_{L,\mathbf{k}\sigma} +(\Delta \hat{c}_{L,\mathbf{k}\uparrow}^\dag \hat{c}_{L,-\mathbf{k}\downarrow}^\dag +{\rm h.c.})\biggr],
\end{align}
where $\xi_k=k^2/2m-\mu$, $\mu$ is the chemical potential, $\Delta$ is the superconducting gap, and $c_{L,\mathbf{k}\sigma}$ ($c^{\dagger}_{L,\mathbf{k}\sigma}$) annihilates (creates) an electron in the tip with momentum $\mathbf{k}$ and spin $\sigma$. The Hamiltonian of the substrate contains a magnetic impurity, located at the origin, with spin $S$ pointing along the $z$ direction. The impurity couples to the substrate via a potential $V\delta(\mathbf{r})$ and exchange coupling $JS\sigma_z\delta(\mathbf{r})$, where $\sigma_z$ is a Pauli matrix in spin space. The Hamiltonian takes the form
\begin{align}
\hat{H}_{R}=\int \frac{d\mathbf{k}}{(2\pi)^3} \biggl[\sum_\sigma \xi_k   \hat{c}_{R,\mathbf{k}\sigma}^\dag  \hat{c}_{R,\mathbf{k}\sigma} +(\Delta \hat{c}_{R,\mathbf{k}\uparrow}^\dag \hat{c}_{R,-\mathbf{k}\downarrow}^\dag +{\rm h.c.})\biggr]+\sum_\sigma (V-JS\sigma) \hat{c}_{R,\sigma}^\dag  \hat{c}_{R,\sigma},\label{app_shiba_hamiltonian}
\end{align}
where the operator $\hat{c}_{R,\sigma}=\int d\mathbf{k}\hat{c}_{R,\mathbf{k}\sigma}/(2\pi)^3$ annihilates an electron with spin $\sigma$ at the origin. One can always choose a gauge such that the superconducting order parameters in tip and substrate are real. The superconducting phase difference $\phi(\tau)$ then enters the tunneling Hamiltonian 
\begin{equation}
\hat{H}_{T}(\tau)=\sum_{\sigma}\left[te^{i\phi(\tau)/2}\hat{c}_{L\sigma}^{\dagger}(\tau)\hat{c}_{R\sigma}(\tau)+te^{-i\phi(\tau)/2}\hat{c}_{R\sigma}^{\dagger}(\tau)\hat{c}_{L\sigma}(\tau)\right],
\end{equation}
where $\tau$ is the time argument, $t$ the hopping strength, and we have made the time dependence of $c_{L/R,\sigma}$ and $c^\dag_{L/R,\sigma}$ explicit. In writing the tunneling Hamiltonian, we have assumed that the substrate is contacted at the impurity location. The time-dependent phase difference between the tip and the sample, $\phi(\tau)=\phi_{0}+2eV\tau$, depends on the voltage $V$ applied to the junction. 

The current operator can be obtained from the Heisenberg equation of motion $\hat{I}=-e\dot{\hat{N}}_{L}=ie[\hat{N}_{L},\hat{H}_{T}]$,
where $\hat{N}_{L}$ is the electron-number operator of the tip. We obtain
\begin{equation}
\hat{I}(\tau)=ie\sum_{\sigma}\left[te^{i\phi(\tau)/2}\hat{c}_{L\sigma}^{\dagger}(\tau)\hat{c}_{R\sigma}(\tau)-te^{-i\phi(\tau)/2}\hat{c}_{R\sigma}^{\dagger}(\tau)\hat{c}_{L\sigma}(\tau)\right].
\end{equation}
Taking the expectation value yields
\begin{equation}
I(\tau)=e\Tr\left\{\tau_{z}\left[\hat{t}(\tau)G_{RL}^{<}(\tau,\tau)-G_{LR}^{<}(\tau,\tau)\hat{t}^{*}(\tau)\right]\right\},
\end{equation}
where $\tau_z$ is a Pauli matrix acting in Nambu space. In the last expression, we introduced the lesser Green function in Nambu space
\begin{align*}
 G_{ij}^{<}(\tau_{1},\tau_{2})=i\begin{pmatrix}\braket{c_{j\uparrow}^\dag(\tau_2) c_{i\uparrow}(\tau_1)} & \braket{c_{j\downarrow}(\tau_2) c_{i\uparrow}(\tau_1)}\\ \braket{c_{j\uparrow}^\dag(\tau_2) c^\dag_{i\downarrow}(\tau_1)} & \braket{c_{j\downarrow}(\tau_2) c^\dag_{i\downarrow}(\tau_1)} \end{pmatrix}
\end{align*}
with $i,j=L,R$ and the hopping matrix 
\begin{equation}
\hat{t}(\tau)=\left(\begin{array}{cc}
te^{i\phi(\tau)/2} & 0\\
0 & -te^{-i\phi(\tau)/2}
\end{array}\right).
\end{equation}
The time dependence only enters the phase difference and thus the current is a periodic function of time $\tau$ with period $2\pi/eV$. We can expand the current in a Fourier series in terms of the frequency $\omega_{0}=eV$
\begin{equation}
I(\tau)=\sum_{m}I_{m}e^{im\omega_{0}\tau}.
\end{equation}
The nonequilibrium Green functions depend on two time arguments and have a generalized Fourier expansion
\begin{equation}
G(\tau_{1},\tau_{2})=\frac{1}{2\pi}\sum_{m}\int d\omega\, e^{-i\omega\tau_{1}}e^{i(\omega+m\omega_{0})\tau_{2}}G(\omega,\omega+n\omega_{0}).
\end{equation}
We adopt the notation $G_{nm}(\omega)=G(\omega+n\omega_{0},\omega+m\omega_{0})$ for which the relation $G_{nm}(\omega)=G_{n-m,0}(\omega+m\omega_{0})$ holds. The hopping matrix and its conjugate are given by
\begin{align}
\hat{t}(\tau)&=\hat{t}^e_{01}e^{i\omega_{0}\tau}+\hat{t}^h_{10}e^{-i\omega_{0}\tau}=\left(\begin{array}{cc}
t & 0\\
0 & 0
\end{array}\right)e^{i\omega_{0}\tau}
+\left(\begin{array}{cc}
0 & 0\\
0 & -t
\end{array}\right)e^{-i\omega_{0}\tau},\\
\hat{t}^*(\tau)&=\hat{t}^h_{01}e^{i\omega_{0}\tau}+\hat{t}^e_{10}e^{-i\omega_{0}\tau}=\left(\begin{array}{cc}
0 & 0\\
0 & -t
\end{array}\right)e^{i\omega_{0}\tau}
+\left(\begin{array}{cc}
t & 0\\
0 & 0
\end{array}\right)e^{-i\omega_{0}\tau},
\end{align}
with $\hat{t}^{e/h}_{nm}=\hat{t}^{e/h}_{n-m,0}$.
Here, we focus on the $dc$ current which is given by the zeroth order in the Fourier expansion,
\begin{align}
I_{0} & =\frac{e}{h}\int d\omega\,\Tr\left[\hat{\sigma}_{z}\left(\hat{t}^e_{01}G_{RL,10}^{<}+\hat{t}^h_{0,-1}G_{RL,-1,0}^{<}-G_{LR,01}^{<}\hat{t}^e_{10}-G_{LR,0,-1}\hat{t}^h_{-1,0}\right)\right]\\
 & =\frac{e}{h}\int d\omega\,\left[tG_{RL,10}^{<,ee}+tG_{RL,-1,0}^{<,hh}-tG_{LR,01}^{<,ee}-tG_{LR,0,-1}^{<,hh}\right],\label{dc_current_green_functions}
\end{align}
where the superscripts $ee$ and $hh$ denote the two diagonal matrix elements in Nambu space.
We do not include the nonresonant Andreev reflections at the superconducting tip and thus neglect multiple Andreev reflection processes, i.e., $g_{L}^{eh}=g_{L}^{he}=0$, where $g_{L}$ denotes the bare Green function of the tip in the absence of the tunnel coupling. Importantly, we retain Andreev reflections at the substrate as they may be resonantly enhanced due to the presence of Shiba bound states. We can now write the Green functions $G_{LR}$ appearing in Eq.~(\ref{dc_current_green_functions}) in terms of $g_L$ and the sample Green functions $G_R$, which includes tunneling only through the self energy of the Shiba state. Using the Langreth rule \cite{Haug2008}
\begin{align}
G_{RL}^{<}&=G_{R}^{r}\hat{t}^{*}g_{L}^{<}+G_{R}^{<}\hat{t}^*g_{L}^{a},\\
G_{LR}^{<}&=g_{L}^{r}\hat{t}G_{R}^{<}+g_{L}^{<}\hat{t}G_{R}^{a},
\end{align}
we obtain
\begin{align}
G_{RL,10}^{<,ee}&\simeq G_{R,11}^{r,ee}\hat{t}_{10}^{e}g_{L,00}^{<ee}+G_{R,11}^{<,ee}\hat{t}_{10}^{e}g_{L,00}^{a,ee},\\
G_{RL,-1,0}^{<,hh}&\simeq G_{R,-1,-1}^{r,hh}\hat{t}_{-1,0}^{h}g_{L,00}^{<,hh}+G_{R,-1,-1}^{<,hh}\hat{t}_{-1,0}^{h}g_{L,00}^{a,hh},\\
G_{LR,01}^{<,ee}&\simeq g_{L,00}^{r,ee}\hat{t}_{01}^eG_{R,11}^{<,ee}+\hat{g}_{L,00}^{<,ee}\hat{t}_{01}^eG_{R,11}^{a,ee},\\
G_{LR,0,-1}^{<,hh}&\simeq g_{L,00}^{r,hh}\hat{t}_{0,-1}^hG_{R,-1,-1}^{<,hh}+g_{L,00}^{<,hh}\hat{t}_{0,-1}^hG_{R,-1,-1}^{a,hh}.
\end{align}
Writing $G(\omega+neV) = G_{nn}(\omega)$, and $g_{L} = g_{L}^{ee}=g_{L}^{hh}$, we obtain for the current 
\begin{align}
I & =\frac{e}{h}t^{2}\int d\omega\,\left\{ \left[G_{R}^{r,ee}(\omega_+)-G_{R}^{a,ee}(\omega_+)\right]g_{L}^{<}(\omega)+G_{R}^{<,ee}(\omega_+)\left[g_{L}^{a}(\omega)-g_{L}^{r}(\omega)\right]\right.\nonumber\\
 & -\left.\left[G_{R}^{r,hh}(\omega_-)-G_{R}^{a,hh}(\omega_-)\right]g_{L}^{<}(\omega)-G_{R}^{<,hh}(\omega_-)\left[g_{L}^{a}(\omega)-g_{L}^{r}(\omega)\right]\right\} ,
\end{align}
where we used the short-hand notation $\omega_\pm=\omega\pm eV$.
By using the relation $G^{<}-G^{>}=G^{a}-G^{r}$, we arrive at
\begin{align}
I=&\frac{e}{h}t^{2}\int d\omega\left\{ G_{R}^{>ee}(\omega)g_{L}^{<}(\omega_-)-G_{R}^{<ee}(\omega)g_{L}^{>}(\omega_-)\right\} \nonumber\\
 &-\frac{e}{h}t^{2}\int d\omega\left\{ G_{R}^{>hh}(\omega)g_{L}^{<}(\omega_+)-G_{R}^{<hh}(\omega)g_{L}^{>}(\omega_+)\right\} .\label{current_formula_green_function}
\end{align}

\subsection{Shiba-bound-state Green function}

To determine the Green function $G_R$ of the substrate, we first calculate the bare Green function $g_R$ neglecting the tunnel coupling to the tip. Without the magnetic impurity, the Green function of a BCS superconductor in Nambu space evaluated at the origin is
\begin{equation}
    g_{R0}(\omega)=-\frac{\pi\nu_{0}}{\sqrt{\Delta^{2}-\omega^{2}}}\left(\begin{array}{cc}
        \omega & \Delta\\
        \Delta & \omega
    \end{array}\right).
\end{equation}
We can include the coupling to the impurity spin in Eq.~(\ref{app_shiba_hamiltonian}) by means of the Dyson equation $g_{R}^{-1}=g_{R0}^{-1}+JS-V\tau_z$, and obtain 
\begin{align}
    g_{R}(\omega)&=\frac{\pi\nu_{0}\sqrt{\Delta^{2}-\omega^{2}}}{(\omega+\alpha\sqrt{\Delta^{2}-\omega^{2}})^{2}-\Delta^{2}-\beta^2(\Delta^2-\omega^2)}
    \left(\begin{array}{cc}
        \omega+(\alpha+\beta)\sqrt{\Delta^{2}-\omega^{2}} & \Delta\\
        \Delta & \omega+(\alpha-\beta)\sqrt{\Delta^{2}-\omega^{2}}
    \end{array}\right)  \nonumber \\
    &= \frac{\pi\nu_0}{2\omega\alpha - (1-\alpha^2+\beta^2)\sqrt{\Delta^2-\omega^2}}
    \left(\begin{array}{cc}
        \omega+(\alpha+\beta)\sqrt{\Delta^{2}-\omega^{2}} & \Delta\\
        \Delta & \omega+(\alpha-\beta)\sqrt{\Delta^{2}-\omega^{2}}
    \end{array}\right), 
    \label{eq:exactGF}
\end{align}
where we introduced the dimensionless parameters $\alpha=\pi\nu_{0}JS>0$ and $\beta= \pi \nu_0 V$. 
The subgap states with $\abs{\omega}<\Delta$ correspond to the poles of the Green function. 
In particular, in our model the Shiba state energy is given by the pole of $g_R$,  
\begin{align}
  \epsilon_{0}=\Delta\frac{1-\alpha^{2}+\beta^2}{\sqrt{(1-\alpha^2+\beta^2)^2+4\alpha^2}}.
\end{align}
To calculate the tunneling into the Shiba state, we only need $g_R(\omega)$ with $\omega$ close to $\epsilon_0$.
In this limit, we set $\omega=\epsilon_0 + \delta\omega$ and expand the denominator in Eq.~(\ref{eq:exactGF}) to linear order in $\delta\omega$, 
\begin{align}
    2\omega\alpha - (1-\alpha^2+\beta^2)\sqrt{\Delta^2-\omega^2}
        &\simeq 2(\epsilon_0+\delta\omega)\alpha 
        - (1-\alpha^2+\beta^2)\sqrt{\Delta^2-\epsilon_0^2}\left( 1-\frac{2\epsilon_0\delta\omega}{\Delta^2- \epsilon_0^2} \right)^{1/2} \nonumber \\
        &=\delta\omega\left(2\alpha + \frac{(1-\alpha^2+\beta^2)\epsilon_0}{\sqrt{\Delta^2-\epsilon_0^2}}\right) \nonumber \\
        &=\delta\omega \frac{(1-\alpha^2+\beta^2)^2+4\alpha^2}{2\alpha}.
\end{align}
The numerator can be evaluated at $\omega=\epsilon_0$, which leads to 
\begin{equation}
    \omega+(\alpha\pm\beta)\sqrt{\Delta^2-\omega^2}\simeq \Delta \frac{1+ (\alpha\pm\beta)^2}{\sqrt{(1-\alpha^2+\beta^2)^2+4\alpha^2}}.
\end{equation}
Thus the Green function has the approximate form
\begin{align}
    g_{R}(\omega)=\frac{1}{\omega-\epsilon_{0}}\left(\begin{array}{cc}
        u^{2} & uv\\
        uv & v^{2}
    \end{array}\right)
\end{align}
with
\begin{equation}
    u^2,v^2 = \frac{2\alpha\pi\nu_0\Delta\left(1+(\alpha\pm\beta)^2\right)}{\left( (1-\alpha^2+\beta^2)^2 + 4\alpha^2\right)^{3/2}}.
\end{equation}
Here, $u$ and $v$ are the electron and hole components of the Shiba state (corresponding to the upper and lower sign, respectively). Note that in general $u\neq v$ when potential scattering by the impurity is included. 

\subsection{Self energy due to relaxation processes}

Phonon or photon induced relaxation processes introduce a self energy $\Sigma_{\rm ph}$ into the substrate Green function
\begin{equation}
   G = g + g\Sigma_{\rm ph} g + g\Sigma_{\rm ph} g\Sigma_{\rm ph} g +\ldots
\end{equation}
Approximating the bare substrate Green function $g$ by the contribution of the Shiba state, 
\begin{equation}
     g(\omega) = |\psi_S\rangle \frac{1}{\omega-\epsilon_0} \langle \psi_S|
\end{equation}
with 
\begin{equation}
    \langle {\bf r} |\psi_S\rangle = \left( \begin{array}{c} u({\bf r}) \\ v({\bf r}) \end{array}\right),
\end{equation}
we find
\begin{equation}
   G(\omega) = |\psi_S\rangle \frac{1}{\omega-\epsilon_0-\langle \psi_S|\Sigma_{\rm ph}(\omega)|\psi_S\rangle } \langle \psi_S|.
\end{equation}
We approximate the self energy by its value at $\omega = \epsilon_0$ and retain only the imaginary part,
\begin{equation}   
 \Gamma_{\rm ph} = 2{\rm Im}  \langle \psi_S|\Sigma_{\rm ph}(\epsilon_0)|\psi_S\rangle. 
\end{equation}
Thus, the retarded and advanced Green functions of the Shiba state read
\begin{equation}
    g_{R}^{r,a}(\omega)=\frac{1}{\omega-\epsilon_{0}\pm i\Gamma_{\rm ph}(\epsilon_0)/2}
    \left(\begin{array}{cc}
        u^{2} & uv\\
        uv & v^{2}
    \end{array}\right).
\end{equation}
Here we have again restricted attention to the Green function at the position of the impurity. 

In quasi-equilibrium, the greater and lesser Green function can be expressed in terms of the retarded and advanced Green functions, 
\begin{align}
    g^{<}_{R}(\omega)&=f(\omega)(g^a_R(\omega)-g^r_R(\omega))
    =\frac{\Sigma_{\rm ph}^{<}(\epsilon_0)}{(\omega-\epsilon_{0})^2+\left(\Gamma_{\rm ph}(\epsilon_0))/2\right)^2}
    \left(\begin{array}{cc}
        u^{2} & uv\\
        uv & v^{2}
    \end{array}\right),\\
    g^{>}_{R}(\omega)&=-(1-f(\omega))(g^a_R-g^r_R)
    =\frac{\Sigma_{\rm ph}^{>}(\epsilon_0)}{(\omega-\epsilon_{0})^2+\left(\Gamma_{\rm ph}(\epsilon_0)/2\right)^2}\left(\begin{array}{cc}
        u^{2} & uv\\
        uv & v^{2}
    \end{array}\right),
\end{align}
where $f(\omega)$ is the quasi-equilibrium distribution function and we used the relations
\begin{equation}
     -i\Sigma_{\rm ph}^{<} = \Gamma_{\rm ph} f\,, \qquad   i\Sigma_{\rm ph}^{>} = \Gamma_{\rm ph}(1-f)\,.\label{self-energy}
\end{equation}
We introduce $\Gamma_1 = i\Sigma^{>}_{ph}(\epsilon_0)$ and $\Gamma_2 = -i\Sigma^{<}_{ph}(\epsilon_0)$ which can be interpreted as the rates with which the Shiba level is emptied or occupied. Note that $\Gamma_{\rm ph} = \Gamma_1 +\Gamma_2$.

\subsection{Self energy due to tip-substrate tunneling}

We now include the tunnel coupling of the Shiba state to the tip. For simplicity, we assume that the tip position is identical with the impurity position. Then, the self energy due to the tunneling is local at the position of the impurity, and we can suppress spatial arguments in the following. The tunneling gives rise to the self energy 
\begin{align}
\Sigma^r_{R}&=\hat{t}_{01}^{h}g^r_{L,11}\hat{t}_{10}^h+\hat{t}_{0-1}^{e}g^r_{L,-1,-1}\hat{t}_{-10}^e 
=t^2
\begin{pmatrix}
 g^r_L(\omega_-)& 0\\
0&  g^r_L(\omega_+) 
\end{pmatrix},
\end{align}
where we neglect Andreev reflections in the tip as discussed in the main text. Similar relations hold for the self energies $\Sigma^{a,<,>}_{R}$. The retarded and advanced Green functions of the Shiba level coupled to the tip can be obtained from the Dyson equation 
\begin{align}
G^{r,a}_R=\frac{1}{1-g_R^{r,a}\Sigma_R^{r,a}} g_R^{r,a} =\frac{1}{\omega-\epsilon_0\pm i\Gamma/2}\left(\begin{array}{cc}
        u^{2} & uv\\
        uv & v^{2}
    \end{array}\right),\label{broadened_shiba_green_function}
\end{align}
where the imaginary part of the self energy leads to a broadening $\Gamma=\Gamma_e(\omega)+\Gamma_h(\omega)+\Gamma_1+\Gamma_2$ with
\begin{align}
\Gamma_e(\omega)&=2\pi t^2 u^2\rho(\omega_-),\\
\Gamma_h(\omega)&=2\pi t^2 v^2\rho(\omega_+) 
\end{align}
in terms of the BCS density of states 
\begin{equation}
    \rho(\omega) =  \nu_0 \frac{\abs{\omega}\theta(\abs{\omega}-\Delta)}{\sqrt{\omega^2 -\Delta^2}}
\end{equation}
with $\nu_0$ the normal density of states at the Fermi energy. In Eq.~(\ref{broadened_shiba_green_function}) we have neglected the real part of the self energy which would lead to a shift of the resonance energy $\propto t^2$.
The lesser Green function of the Shiba state is given by \cite{Haug2008} 
\begin{align}
 G_R^<&=g_R^<+g_R^r\Sigma_R^rG_R^<+g_R^r\Sigma_R^<G_R^a+g_R^<\Sigma_R^aG_R^a\\
 &=\frac{1}{1-g_R^r\Sigma_R^r}[g_R^<(1+\Sigma_R^aG_R^a)+g_R^r\Sigma_R^<G_R^a]\,.
\end{align}
A straightforward calculation using the relations $g_{L}^{<}(\omega)=2\pi i\rho(\omega)n_F(\omega)$ and $g_{L}^{>}(\omega)=-2\pi i\rho(\omega)(1-n_F(\omega))$ reveals
\begin{align}
 G_R^<&= i\frac{ \Gamma_2+\Gamma_e(\omega)n_F(\omega_-)+\Gamma_h(\omega)n_F(\omega_+)}{(\omega-\epsilon_0)^2+(\Gamma_1+\Gamma_2+\Gamma_e(\omega)+\Gamma_h(\omega))^2/4}\begin{pmatrix}u^2& uv\\ uv& v^2\end{pmatrix}.
 \end{align}
Along the same lines we find
\begin{align}
 G_R^>&= -i\frac{\Gamma_1+\Gamma_e(\omega)(1-n_F(\omega_-))+\Gamma_h(\omega)(1-n_F(\omega_+))}{(\omega-\epsilon_0)^2+(\Gamma_1+\Gamma_2+\Gamma_e(\omega)+\Gamma_h(\omega))^2/4}\begin{pmatrix}u^2& uv\\ uv& v^2\end{pmatrix}.
 \end{align}

\subsection{Expressions for the tunneling current}

We can now evaluate the current in Eq.~(\ref{current_formula_green_function}) which yields $I=I^s+I^a$, where
\begin{align}
I^s(V)&=\frac{e}{h}\int d\omega
\frac{\Gamma_{1}[\Gamma_{e}(\omega)n_F(\omega_-)-\Gamma_{h}(\omega)n_F(\omega_+)]-\Gamma_{2}[\Gamma_{e}(\omega)(1-n_F(\omega_-))-\Gamma_{h}(\omega)(1-n_F(\omega_+))]
}{(\omega-\epsilon_{0})^{2}+\left(\Gamma(\omega)/2\right)^{2}},\label{IR_app}\\
I^a(V)&=\frac{2e}{h}\int d\omega\frac{\Gamma_{e}(\omega)\Gamma_{h}(\omega)}{(\omega-\epsilon_{0})^{2}+\left(\Gamma(\omega)/2\right)^{2}}\left[n_F(\omega_-)-n_F(\omega_+)\right].\label{IA_app}
\end{align}
The current $I^a$ originates from resonant Andreev reflection, whereas $I^s$ describes single-particle tunneling and subsequent relaxation of quasiparticles in the Shiba state. These two equations are given in the main text as Eqs.~(1)
and (2).

\subsection{Relaxation processes}\label{sec:relaxation}

The intrinsic broadening of the Shiba level is determined by the rates $\Gamma_{1}$ for emptying and $\Gamma_2$ for filling a Shiba state. In Ref.~\cite{martin13} the authors considered elastic processes due to a coupling to a fermionic bath as a source of quasiparticle relaxation. 
Another possible relaxation mechanism involves transitions between the Shiba state and the quasiparticle continuum assisted by phonons or photons.

In the present experiment, we find no indication for the presence of a quasiparticle bath at subgap energies as the Pb sample exhibits a hard gap away from the impurities. A more likely source are phonon-assisted transitions which result in a thermal distribution of the Shiba state $f=n_F(\epsilon_0)$ in the absence of a tunnel coupling, where $f=\Gamma_2/(\Gamma_1+\Gamma_2)$ [see Eq.~(\ref{self-energy})].
Thus for purely thermal relaxation we generally find 
\begin{align}
\frac{\Gamma_1}{\Gamma_2}=e^{\epsilon_0/T}. \label{detailed_balance}
\end{align}
The most basic relaxation process involves direct transitions between the Shiba state and the quasiparticle continuum.
Such processes were studied in Ref.~\cite{kozorezov} where it was shown that the relaxation rates are given by 
\begin{align}
\Gamma_{1}&\sim\sqrt{\frac{T}{\Delta}}e^{-(\Delta-\epsilon_0)/T}\left[(\Delta-\epsilon_{0})+(\Delta+\epsilon_{0})e^{-\epsilon_{0}/T}\right]\label{gamma1_temp},\\
\Gamma_{2}&\sim\sqrt{\frac{T}{\Delta}}e^{-\Delta/T}\left[(\Delta-\epsilon_{0})+(\Delta+\epsilon_{0})e^{-\epsilon_{0}/T}\right].
\end{align}
The relaxation rate $\Gamma_1$ for leaving the Shiba state has a thermal factor $\exp[-(\Delta-\epsilon_0)/T]$ involving the required phonon energy of the transition to the continuum $\Delta-\epsilon_0$, whereas $\Gamma_2$ is limited by the thermal occupation $\exp(-\Delta/T)$ of the excited quasiparticles in the continuum. The ratio of the two rates indeed yields Eq.~(\ref{detailed_balance}).

We study the nature of the relaxation processes in the experiment by determining the relaxation rates at different temperatures (see discussion in main text and Sec.~\ref{sec:lifetime}). While our findings are generally consistent with thermal relaxation, the data cannot be readily explained in terms of direct transitions to the continuum. Instead, we propose a cascade of thermal relaxation processes via intermediate Shiba states as an alternative mechanism consistent with our experimental findings.

\section{Calculation of current and differential conductance at the thresholds}

In this section, we provide details of the calculations underlying Eqs.\ (3--5) for the peak conductances in the main text. We also calculate the currents at these bias voltages which were used in the main text to extract relaxation rates. At the end of this section we illustrate these results by numerically calculating the current and the differential conductance as a function of tunneling strength. We organize the calculation by threshold voltages.

\subsection{\texorpdfstring{\protect{$eV=\Delta+\epsilon_0$}}{}}

At this threshold, there are two contributing processes to the current, namely single-electron tunneling into the Shiba state as well as resonant Andreev processes: $I_{\Delta+\epsilon_0}=I^s_{\Delta+\epsilon_0}+I^a_{\Delta+\epsilon_0}$. We find 
\begin{equation}
   I_{\Delta+\epsilon_0} = \left\{\begin{array}{ccc} \frac{2e}{\hbar} (2\Gamma_h+\Gamma_1) \frac{\omega_e^{3/2}}{\Gamma_1^{3/2}}& & \omega_{e} \ll \Gamma_1 \\
   \frac{e}{3\hbar}(2\Gamma_h +\Gamma_1) & & \omega_{e} \gg \Gamma_1
   \end{array}\right.
\label{I1}
\end{equation}
for the peak current and
\begin{equation}
   \alpha_+ = \left\{\begin{array}{ccc} \frac{4\pi e^2}{h} \frac{2\Gamma_h+\Gamma_1}{\Gamma_1} \frac{\omega_{e}^{3/2}}{\Gamma_1^{3/2}}& &    
   \omega_{e} \ll \Gamma_1 \\
   \frac{4\pi e^2}{9h} \frac{2\Gamma_h +\Gamma_1}{\omega_{e}} & & \omega_{e} \gg \Gamma_1
   \end{array}\right.
\label{alphaplus}
\end{equation}
for the peak conductance. We will now derive these results, first treating the single-electron processes and subsequently analyzing the Andreev process. The quantities entering into these expressions will be defined as the calculation proceeds. 

\subsubsection{Single-electron tunneling}

The relevant single-electron process is shown in Fig.\ 2(a) in the main text. Analytically, this process contributes the term
\begin{equation}
  I^s_{\Delta+\epsilon_0}(V) = e\int \frac{d\omega}{2\pi\hbar} \frac{\Gamma_1\Gamma_e(\omega)n_F(\omega-eV)}{(\omega-\epsilon_0)^2+(\Gamma/2)^2}\label{app_current_relaxation}
\end{equation}
to the current. Here, the subscript denotes the threshold and the superscript the single-particle (as opposed to Andreev) nature of the contributions. The integration variable $\omega$ denotes the energy of the tunneling electron as measured from the Fermi energy of the substrate superconductor. In Eq.~(\ref{app_current_relaxation}) and throughout this section we focus on contributions to the current which originate from the vicinity of the BCS singularity. Unlike the differential conductance, the full current at the main thresholds $e|V|=\Delta+\epsilon_0$ also includes the contributions from terms $\propto \Gamma_2$ in Eq.~(\ref{IR_app}), which are responsible for the thermal peaks. The quantitative comparison with experiment in Sec.~\ref{sec:theo_fit} includes all contributions to the current.

At the threshold $eV=\Delta+\epsilon_0$, the coupling $\Gamma_e(\omega)$ becomes singular exactly at the Shiba energy $\omega=\epsilon_0$ because of the diverging BCS density of states $\rho(\omega-eV)$ [cf.\ Fig.\ 2(a)]. The dominant contribution to the current comes from the vicinity of the singularity at $\omega=eV-\Delta\sim \epsilon_0$ and we can approximate 
\begin{equation}
  \Gamma_e(\omega) =2\pi u^2t^2\rho(\omega-eV)\simeq \gamma_e \sqrt{\frac{\Delta}{2}}\, \frac{\theta(eV-\Delta-\omega)}{\sqrt{eV-\Delta-\omega}}
\end{equation} 
in terms of the normal-state tunneling rate $\gamma_e=2\pi u^2\nu_0t^2$. 
In the region of interest the thermal occupation of the tip is $n_F(\omega-eV)\simeq 1$. Note that this insensitivity to thermal smearing is a consequence of the superconducting tip. (Of course, the current is still sensitive to temperature which enters into the relaxation rates $\Gamma_1$ and $\Gamma_2$.) This also implies that the bias voltage enters into the current only via the BCS density of states which is quite distinct from the case of a normal-state tip. 

With these ingredients, we can now compute the current in the vicinity of the threshold,
\begin{equation}
  I^s_{\Delta+\epsilon_0}(V) \simeq e \gamma_e\Gamma_1 \sqrt{\frac{\Delta}{2}} \int \frac{d\omega}{2\pi\hbar}   
  \frac{1}{(\omega-\epsilon_0)^2+\left(\frac{\Gamma_1}{2}+\frac{\gamma_e\sqrt{\Delta/2}}{2\sqrt{eV-\Delta-\omega}}\right)^2}
  \frac{\theta(eV-\Delta-\omega)}{\sqrt{eV-\Delta-\omega}}.
\end{equation}
Here, we used that $\Gamma_1\gg\Gamma_2$ for $\epsilon_0\gg T$ and $\Gamma_h \ll \Gamma_e$. The latter will be justified below. Note that it is however important to keep both $\Gamma_1$ and $\Gamma_e$. 

\begin{figure}[t]
\centering
\includegraphics[width=.35\textwidth]{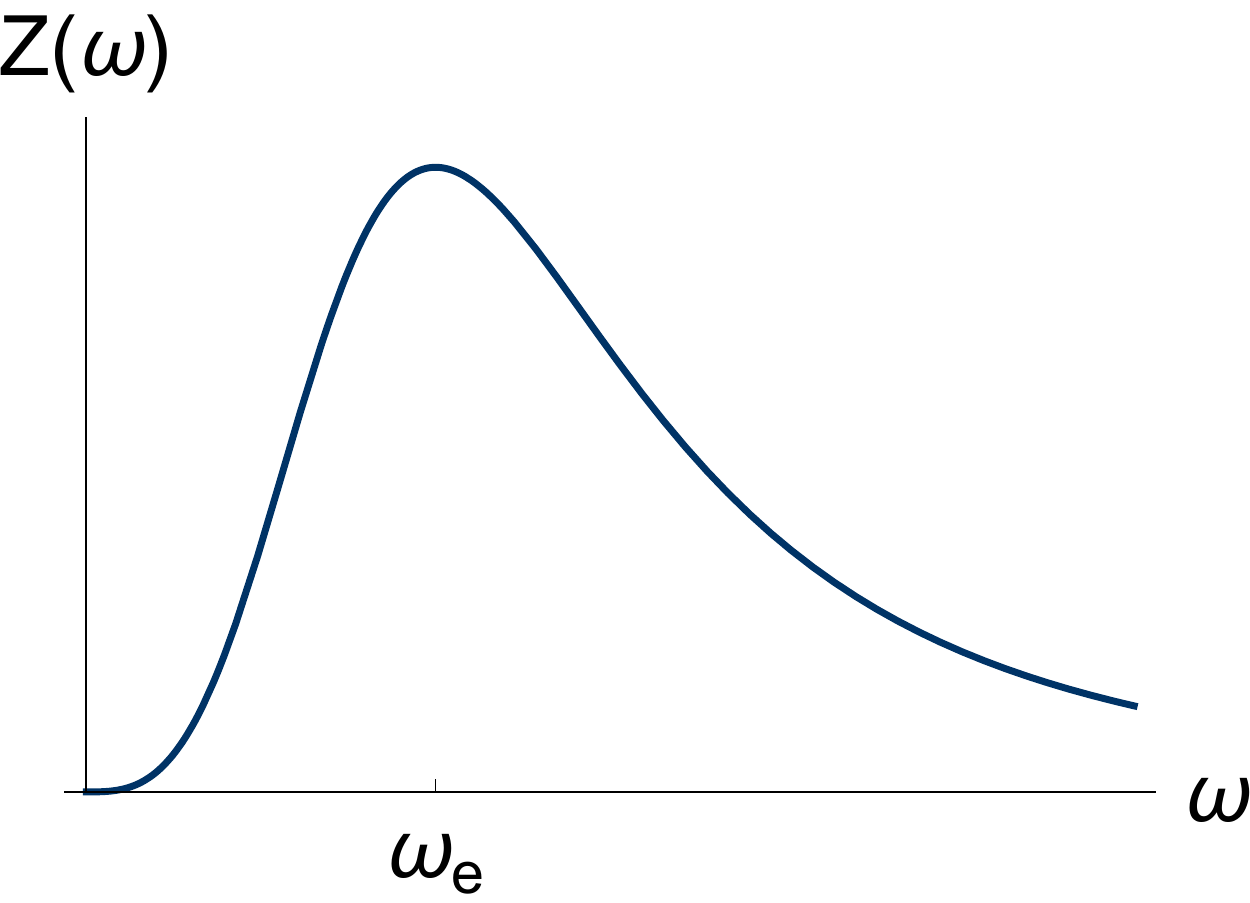}
\caption{Characteristic function $Z(\omega)$ defined in the text at the threshold ($eV'=0$) and for $\Gamma_1=0$. This function determines the conductance via Eq.~(\ref{conductance_integral}). It is zero at the Shiba state ($\omega=0$) and peaks at $\omega_e$.}
\label{fig:spec_func}
\end{figure}

We simplify notation by measuring voltages from the threshold,
\begin{equation}
   eV' = eV-(\Delta+\epsilon_0),
\end{equation}
and introducing the characteristic energy
\begin{align}
   \omega_{e} &= \left(\frac{1}{2} \gamma_e \sqrt{\frac{\Delta}{2}}\right)^{2/3}.
\end{align}
Then, we have
\begin{equation}
  I^s_{\Delta+\epsilon_0}(V') \simeq 2e \omega_e^{3/2}\Gamma_1  \int_0^\infty \frac{d\omega}{2\pi\hbar} \frac{1}{\sqrt{\omega}}  
  \frac{1}{(\omega-eV')^2+\left(\frac{\Gamma_1}{2}+\frac{\omega_e^{3/2}}{\sqrt{\omega}}\right)^2}
\end{equation}
and
\begin{equation}
  G^s_{\Delta+\epsilon_0}(V') \simeq 4 e^2  \omega_e^{3/2}\Gamma_1  \int_0^\infty \frac{d\omega}{2\pi\hbar} \frac{1}{\sqrt{\omega}}  
  \frac{\omega-eV'}{\left[(\omega-eV')^2+\left(\frac{\Gamma_1}{2}+\frac{\omega_e^{3/2}}{\sqrt{\omega}}\right)^2\right]^2}\label{conductance_integral}
\end{equation}
for the corresponding conductance $G=dI/dV$. The conductance involves an integral over the tip density of states $\sim 1/\sqrt{\omega}$ multiplied by the function $Z(\omega)=(\omega-eV')/[(\omega-eV')^2+(\Gamma_1/2+\omega_e^{3/2}/\sqrt{\omega})^2]^2$.
We emphasize that $Z(\omega)$ is not the spectral function of the Shiba state as it would be for a normal metal tip. This special feature of the superconducting tip arises because the voltage dependence enters through the tip density of states rather than the occupation numbers. As shown in Fig.~\ref{fig:spec_func} $Z(\omega)$ vanishes at the Shiba state (now at $\omega=0$) because of the divergent broadening induced by the superconducting tip and also vanishes far from the Shiba energy. In between, it peaks at a scale set by the maximum of the effective tunneling strength $\omega_e$ and the thermal relaxation rate $\Gamma_1$ (in Fig.~\ref{fig:spec_func} we have set $\Gamma_1=0$ in which case the peak is at $\omega_e$).

We can now also discuss the hole tunneling rate $\Gamma_h(\omega)$ with an associated energy scale $   \omega_{h} = ( \gamma_h \sqrt{\Delta})^{2/3}/2$ in terms of the normal state hole tunneling rate $\gamma_h=2\pi v^2\nu_0t^2$. In principle, hole tunneling introduces another term into the broadening of $Z(\omega)$. The broadening then becomes
\begin{equation}
  \frac{\Gamma_1}{2}+\frac{\omega_e^{3/2}}{\sqrt{\omega}}+\frac{\omega_h^{3/2}}{\sqrt{\omega+2\epsilon_0}}.
\end{equation}
We can neglect the last term (i.e., the hole contribution to the width of $Z(\omega)$) as long as ${\rm max}\{\Gamma_1,\omega_{e}\}\gg \sqrt{\omega_h^{3}/\epsilon_0}$. 
In principle, one may imagine situations in which $v\gg u$ so that $\Gamma_h(\omega)$ contributes significantly to the broadening in the strong tunneling regime. As this case is probably irrelevant for this experiment we exclude it from our analytical considerations. We discuss implications of a broadening due to $\Gamma_h(\omega)$ in the presentation of the numerical results at the end of this section.

We focus attention on the peak current and peak conductance. The peak occurs approximately at the threshold bias $eV=\Delta+\epsilon_0$ and we restrict our analytical considerations to the threshold, setting $eV'=0$ in the following. While this makes our analysis more transparent it also introduces a small numerical error. We emphasize that our results exhibit the correct parametric dependence and the quantitative analysis in Sec.~\ref{sec:theo_fit} is based on the numerically exact peak heights. The peak position and height relative to the threshold are discussed in detail in Sec.~\ref{sec:app_lineshape}.

Evaluated at the threshold, the integral for $I^s_{\Delta+\epsilon_0}$ contains the two energy scales $\Gamma_1$ and $\omega_{e}$. For weak tip-substrate tunneling, $\omega_{e}\ll \Gamma_1$, we can neglect the contribution of $\Gamma_e$ to the broadening of $Z(\omega)$. In this limit, we find
\begin{eqnarray}
  I^s_{\Delta+\epsilon_0} &\simeq& 2e  \omega_e^{3/2}\Gamma_1  \int_0^\infty \frac{d\omega}{2\pi\hbar} \frac{1}{\sqrt{\omega}}  
  \frac{1}{\omega^2+{\Gamma^2_1}/{4}} \nonumber\\
      &=& \frac{2\sqrt{2} e}{\pi\hbar}  \frac{ \omega_e^{3/2}}{(\Gamma_1)^{1/2}}  \int_0^\infty \frac{dx}{\sqrt{x}}   \frac{1}{x^2+1}. 
\end{eqnarray}
The integral is elementary and we obtain the result
\begin{eqnarray}
  I^s_{\Delta+\epsilon_0}\simeq \frac{2e}{\hbar} \frac{ \omega_e^{3/2}}{(\Gamma_1)^{1/2}}. 
\end{eqnarray}
The peak conductance in the regime $\omega_{e}\ll \Gamma_1$ can be calculated along the same lines,
\begin{eqnarray}
  G^s_{\Delta+\epsilon_0} &\simeq& \frac{4 e^2}{h} \omega_e^{3/2}\Gamma_1  \int_0^\infty {d\omega}   
  \frac{\sqrt{\omega}}{\left[\omega^2+{\Gamma_1^2}/{4}\right]^2} \nonumber\\
   &=& \frac{16\sqrt{2} e^2}{h} \frac{\omega_e^{3/2}}{(\Gamma_1)^{3/2}}  \int_0^\infty dx   
  \frac{\sqrt{x}}{[x^2+1]^2} \nonumber\\
   &=& \frac{4\pi e^2}{h} \frac{ \omega_e^{3/2}}{(\Gamma_1)^{3/2}},  
\end{eqnarray}
where the $x$-integration is again elementary. 

In the opposite limit of strong tip-substrate tunneling, $\omega_{e}\gg \Gamma_1$, we can neglect the contribution of $\Gamma_1$ to the broadening of $Z(\omega)$. In this limit, we find
\begin{eqnarray}
  I^s_{\Delta+\epsilon_0} &\simeq& 2e  \omega_e^{3/2}\Gamma_1  \int_0^\infty \frac{d\omega}{2\pi\hbar} \frac{1}{\sqrt{\omega}}  
  \frac{1}{\omega^2+\omega_e^3/\omega} \nonumber\\
      &=& \frac{e}{\pi\hbar} \Gamma_1  \int_0^\infty \frac{dx}{\sqrt{x}}   \frac{1}{x^2+1/x}. 
\end{eqnarray}
Performing the integral yields
\begin{eqnarray}
  I^s_{\Delta+\epsilon_0}\simeq \frac{e\Gamma_1}{3\hbar}. 
\end{eqnarray}
The peak conductance in the regime $\omega_{e}\gg \Gamma_1$ can be calculated along the same lines,
\begin{eqnarray}
  G^s_{\Delta+\epsilon_0} &\simeq& \frac{4e^2}{h}  \omega_e^{3/2}\Gamma_1  \int_0^\infty d\omega   
  \frac{\sqrt{\omega}}{\left[\omega^2+\omega_e^3/\omega\right]^2} \nonumber\\
      &=& \frac{4e^2}{h} \frac{\Gamma_1}{\omega_e}  \int_0^\infty dx   \frac{\sqrt{x}}{[x^2+1/x]^2} \nonumber\\
      &=& \frac{4\pi e^2}{9h} \frac{\Gamma_1}{\omega_e}. 
\end{eqnarray}

\subsubsection{Andreev contribution}

The Andreev current is given by
\begin{equation}
  I^a_{\Delta+\epsilon_0}(V) = 2e\int \frac{d\omega}{2\pi\hbar} \frac{\Gamma_e(\omega)\Gamma_h(\omega)[n_F(\omega-eV)-n_F(\omega+eV)]}{(\omega-\epsilon_0)^2+(\Gamma/2)^2}.
\end{equation}
The Fermi functions can be approximated by $n_F(\omega -eV)\simeq 1$ and $n_F(\omega + eV)\simeq 0$, since $\omega \sim \epsilon_0$ and $eV\sim \Delta+\epsilon$ and with $\omega_{e}\ll \epsilon_0$ we can approximate  $\Gamma_h(\omega)$ by a constant. In the case $\epsilon_0\ll\Delta$ it simply reads
\begin{equation}
   \Gamma_h(\omega)\simeq\Gamma_h =  \sqrt{\frac{2\omega_{h}^{3}}{\epsilon_0}}.
\end{equation} 
With these approximations, the integrals become equal to those for the single-particle contribution, with the replacement $\Gamma_1\to\Gamma_h$ in the numerator and an overall prefactor of two. Note that we can still ignore $\Gamma_h$ in the broadening of $Z(\omega)$ under the assumptions spelled out above. This yields the result summarized in Eqs.\ (\ref{I1}) and (\ref{alphaplus}) above.

\begin{figure}[t]
\centering
\includegraphics[width=.2\textwidth]{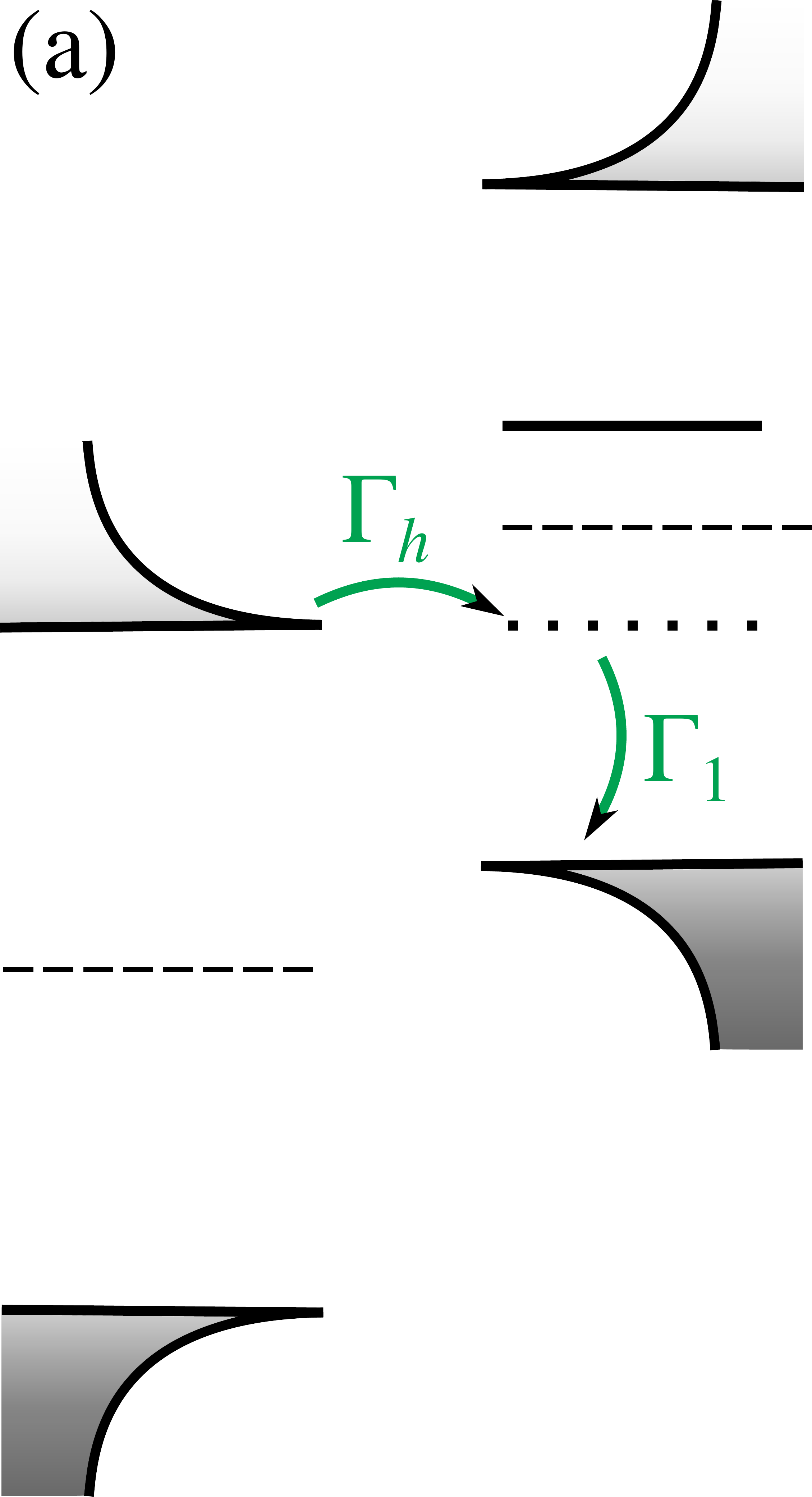}
\hspace{1em}
\includegraphics[width=.2\textwidth]{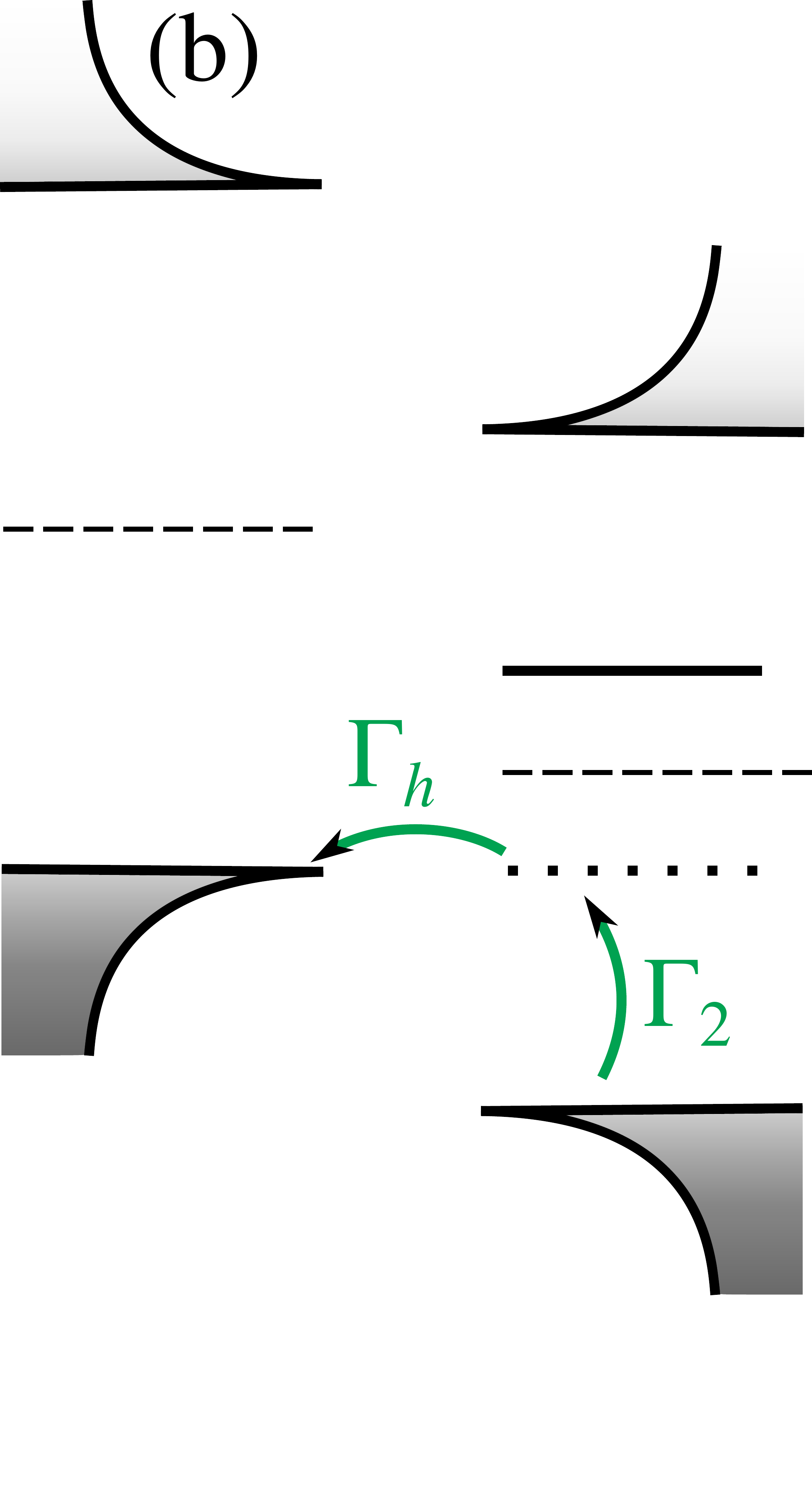}
\caption{Single-particle tunneling processes at (a) $eV=-(\Delta+\epsilon_0$) and (b) $eV=\Delta-\epsilon_0$. At the threshold in (a), an additional Andreev process contributes to the current (cf.\ Fig.~2(b) of the main text).}
\label{fig:tunnel_process_v}
\end{figure}

\subsection{\texorpdfstring{\protect{$eV=-(\Delta+\epsilon_0)$}}{}}

At threshold the current is again the sum of single-electron and Andreev processes. The relevant single-electron process is shown in Fig.~\ref{fig:tunnel_process_v}. Analytically, this process contributes the term
\begin{equation}
  I^s_{-\Delta-\epsilon_0} (V)= -e\int \frac{d\omega}{2\pi\hbar} \frac{\Gamma_1\Gamma_h(\omega)n_F(\omega+eV)}{(\omega-\epsilon_0)^2+(\Gamma/2)^2}
\end{equation}
to the current. Up to overall signs, this differs from the corresponding process near $eV=\Delta+\epsilon_0$ discussed above by exchanging the roles of $\Gamma_e(\omega)$ and $\Gamma_h(\omega)$. The same exchange characterizes the Andreev contribution. Thus, in effect, we can obtain the results for this threshold by interchanging $u\leftrightarrow v$ in the expressions for $eV=\Delta+\epsilon_0$. This yields 
\begin{equation}
   I_{-\Delta-\epsilon_0} = \left\{\begin{array}{ccc} \frac{2e}{\hbar} (2\Gamma_h+\Gamma_1) \frac{\omega_{h}^{3/2}}{\Gamma_1^{3/2}}& & \omega_{h} \ll \Gamma_1 \\
   \frac{e}{3\hbar}(2\Gamma_h +\Gamma_1) & & \omega_{h} \gg \Gamma_1
   \end{array}\right.
\label{I2}
\end{equation}
for the peak current and
\begin{equation}
   \alpha_- = \left\{\begin{array}{ccc} \frac{4\pi e^2}{h} \frac{2\Gamma_h+\Gamma_1}{\Gamma_1} \frac{\omega_{h}^{3/2}}{\Gamma_1^{3/2}}& &    
   \omega_{h} \ll \Gamma_1 \\
   \frac{4\pi e^2}{9h} \frac{2\Gamma_h +\Gamma_1}{\omega_{h}} & & \omega_{h} \gg \Gamma_1
   \end{array}\right.
\label{alphaminus}
\end{equation}
for the peak conductance.

\subsection{\texorpdfstring{\protect{$eV=-(\Delta-\epsilon_0)$}}{}}

At this thermal threshold, only single-electron processes contribute which are shown in Fig.\ 2(c) 
in the main text. Analytically, this process is described by
\begin{equation}
  I^s_{-\Delta+\epsilon_0} (V)= -e\int \frac{d\omega}{2\pi\hbar} \frac{\Gamma_2\Gamma_e(\omega)[1-n_F(\omega-eV)]}{(\omega-\epsilon_0)^2+(\Gamma/2)^2}\,.
\end{equation}
Noting that 
\begin{equation}
   1-n_F(\omega-eV) \simeq 1 = 1 - n_F(e|V|+\omega) \simeq 1,
\end{equation}
we see that this differs from the expression for the single-electron current at the threshold $eV=\Delta+\epsilon_0$ merely by a factor $\Gamma_2/\Gamma_1$. Thus, we obtain
\begin{equation}
   I_{-\Delta+\epsilon_0} = \left\{\begin{array}{ccc} \frac{2e}{\hbar} \Gamma_2 \frac{\omega_{e}^{3/2}}{\Gamma_1^{3/2}}& & \omega_{e} \ll \Gamma_1 \\
   \frac{e}{3\hbar}\Gamma_2 & & \omega_{e} \gg \Gamma_1
   \end{array}\right.
\label{I3}
\end{equation}
for the peak current and
\begin{equation}
   \beta_- = \left\{\begin{array}{ccc} \frac{4\pi e^2}{h} \frac{\Gamma_2}{\Gamma_1} \frac{\omega_{e}^{3/2}}{\Gamma_1^{3/2}}& &    
   \omega_{e} \ll \Gamma_1 \\
   \frac{4\pi e^2}{9h} \frac{\Gamma_2}{\omega_{e}} & & \omega_{e} \gg \Gamma_1
   \end{array}\right.
\label{betaminus}
\end{equation}
for the peak conductance.

\subsection{\texorpdfstring{\protect{$eV=\Delta-\epsilon_0$}}{}}

This thermal threshold is dominated by the contribution shown in Fig.~\ref{fig:tunnel_process_v} and given by
\begin{equation}
  I^s_{\Delta-\epsilon_0} (V)= e\int \frac{d\omega}{2\pi\hbar} \frac{\Gamma_2\Gamma_h(\omega)[1-n_F(\omega+eV)]}{(\omega-\epsilon_0)^2+(\Gamma/2)^2}\,.
\end{equation}
This differs from the thermal threshold at $eV=-\Delta+\epsilon_0$ by the replacement of $\Gamma_e(\omega)$ by $\Gamma_h(\omega)$. Thus, we obtain the current at this threshold by the replacement $u\leftrightarrow v$. This yields
\begin{equation}
   I_{\Delta-\epsilon_0} = \left\{\begin{array}{ccc} \frac{2e}{\hbar} \Gamma_2 \frac{\omega_{h}^{3/2}}{\Gamma_1^{3/2}}& & \omega_{\rm 
   th} \ll \Gamma_1 \\
   \frac{e}{3\hbar}\Gamma_2 & & \omega_{h} \gg \Gamma_1
   \end{array}\right.
\label{I4}
\end{equation}
for the peak current and
\begin{equation}
   \beta_+ = \left\{\begin{array}{ccc} \frac{4\pi e^2}{h} \frac{\Gamma_2}{\Gamma_1} \frac{\omega_{h}^{3/2}}{\Gamma_1^{3/2}}& &    
   \omega_{h} \ll \Gamma_1 \\
   \frac{4\pi e^2}{9h} \frac{\Gamma_2}{\omega_{h}} & & \omega_{h} \gg \Gamma_1
   \end{array}\right.
\label{betaplus}
\end{equation}
for the peak conductance.

\subsection{Discussion and numerical results}\label{sec:app_discussion}

\begin{figure}[t!]
\centering
\includegraphics[width=.46\textwidth]{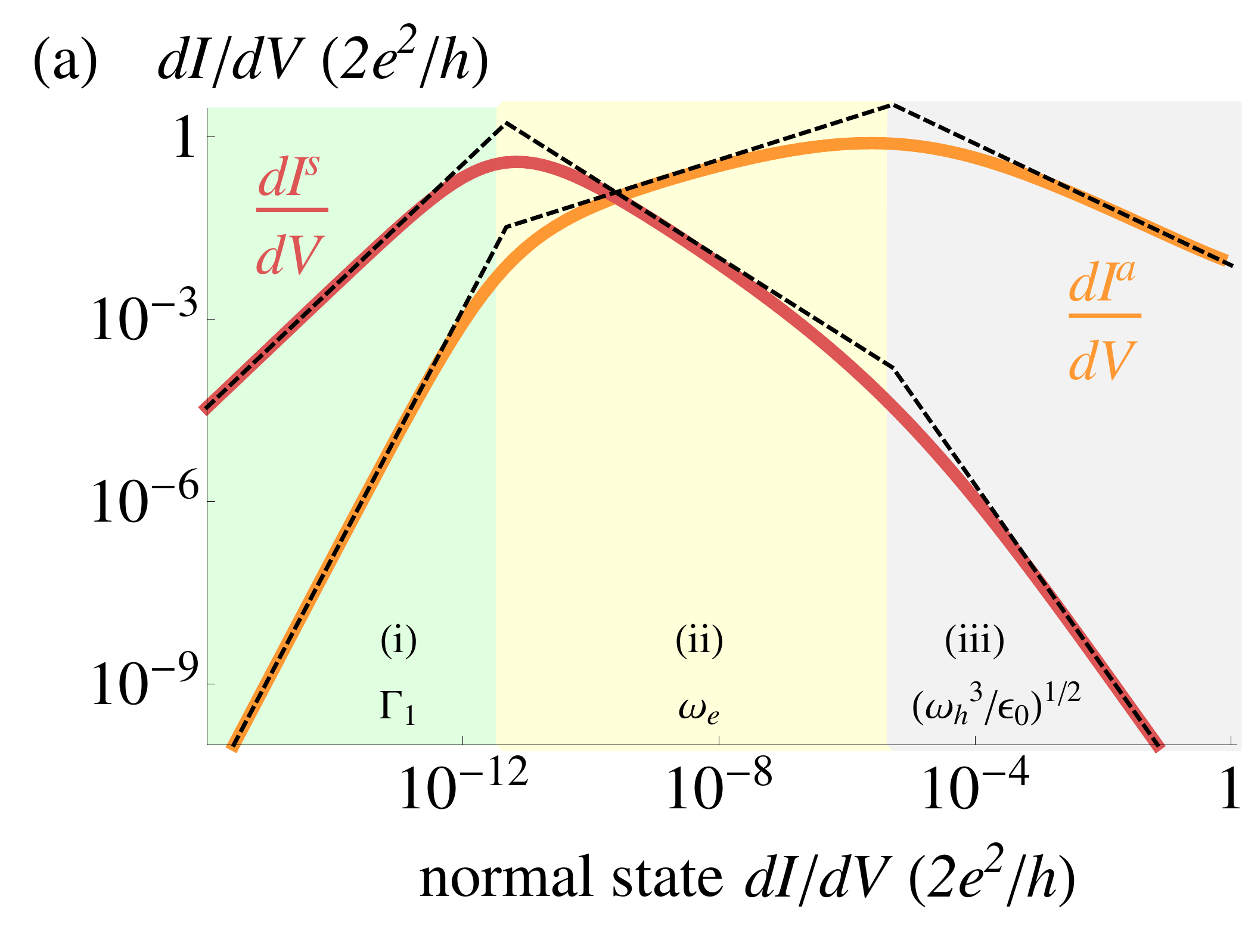}
\includegraphics[width=.46\textwidth]{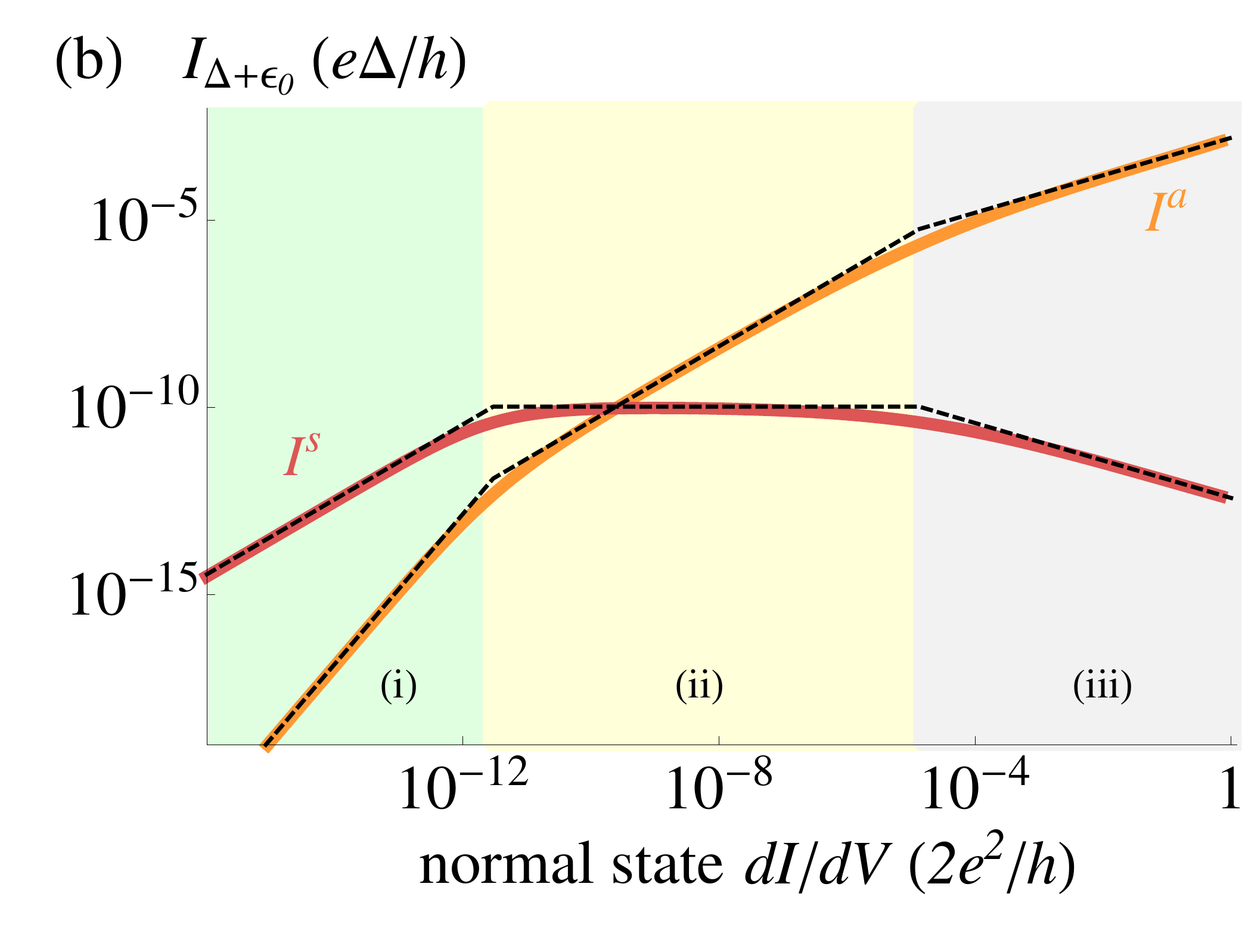}
\caption{(a) Differential conductance and (b) current at the threshold $eV=\Delta+\epsilon$ {\em vs.} normal state conductance. The curves are obtained numerically from Eqs.~(1) and (2) of the main text. The single-particle (red) and Andreev (orange) contributions have maxima at distinct values of normal state $dI/dV$ which separate three regimes with dominating relaxation mechanisms $\Gamma_1$, $\omega_e$, and $(\omega_h^3/\epsilon_0)^{1/2}$. While analytical expressions for the asymptotes (dashed lines) in the first two regimes are given by Eqs.~(\ref{I1}) and (\ref{alphaplus}), a similar analysis also yields expressions in regime (iii). The parameters are chosen such that all three regimes are visible. We have set $\Gamma_1=10^{-10}$, $\Gamma_2=0$, $u^2/\nu_0=0.001$, $v^2/\nu_0=1$, and $\epsilon_0=0.3$, where all energies are measured in units of $\Delta$.}
\label{fig:double_peak_theo}
\end{figure}

In Fig.~\ref{fig:double_peak_theo}(a) we plot the differential conductance from the single-particle and Andreev currents at the threshold $eV=\Delta+\epsilon_0$ according to Eqs.~(\ref{IR_app}) and (\ref{IA_app}) together with the analytical expression in Eq.~(\ref{alphaplus}). We identify three regimes as a function of the normal state conductance $G_N\sim (2e^2/h)t^2\nu_0^2$ which exhibit characteristic power-laws as a function of tunneling strength. These regimes can be associated with different dominant broadening mechanisms (from weak to strong tunneling): (i) intrinsic relaxation $\Gamma_1$, (ii) electron tunneling $\omega_e$, and (iii) hole tunneling $(\omega_h^3/\epsilon_0)^{1/2}$.

The crossover between regimes (i) and (ii) occurs at a normal state conductance $G^1_N$, which can be evaluated from the condition $\Gamma_1\sim \omega_e$. Equivalently the second crossover point $G_N^2$ is obtained from $\omega_e\sim(\omega_h^3/\epsilon_0)^{1/2}$ and we find
\begin{align}\
G_N^{1}&\sim \frac{2e^2}{h} \frac{\nu_0\Delta}{ u^2} \left(\frac{\Gamma_1}{\Delta}\right)^{3/2}\label{Gn1},\\
G_N^{2}&\sim \frac{2e^2}{h}\frac{ u^4\nu_0\Delta}{v^6}\left(\frac{\epsilon_0}{\Delta}\right)^{3/2}\label{Gn2}.
\end{align}

Figure \ref{fig:double_peak_theo}(a) shows that the full differential conductance peak $\alpha_+=dI^s/dV+dI^a/dV$ consists of the sum of two terms that peak at different tunneling strengths and thus typically exhibits two peaks as a function of normal state conductance. At the first crossover point the single-particle contribution $dI^s/dV$ reaches a maximum of order $2e^2/h$. This is readily understood from the single-particle current in Eq.~(\ref{IR_app}), which can be viewed as a resonant tunneling process through the Shiba state with rates $\Gamma_e(\omega)$ and $\Gamma_1$ as depicted in Fig.~2(a) of the main text. When the effective electron tunneling rate $\omega_e$ is equal to $\Gamma_1$ the conductance reaches a universal value of the order of the conductance quantum. For stronger couplings the conductance decreases and the single-particle current shown in Fig.~\ref{fig:double_peak_theo}(b) saturates to a value determined by the relaxation rate. Indeed, Eq.~(\ref{I1}) yields a current $I^s= e\Gamma_1/3\hbar$ independent of tunneling strength in this regime. At even stronger coupling the Andreev current $I^a$ exceeds the single-particle contribution and thus the total current, $I^s+I^a$, exhibits a shoulder as a function of tunnel coupling. 
We remind the reader that the current at the main thresholds would have additional contributions from the terms $\propto\Gamma_2\Gamma_{e/h}(\omega)$ in Eq.~(\ref{IR_app}), which we have excluded from our analytical considerations. We evaluate the full current for the quantitative comparison between theory of experiment in Sec.~\ref{sec:theo_fit}.

We can estimate the normal state conductance $G_N^{*}$ at which the Andreev current becomes the dominant contribution to the current at the threshold $eV=\Delta+\epsilon_0$ from $\Gamma_1\sim\Gamma_h$. We obtain
\begin{align}
 G_N^{*}\sim \frac{2e^2}{h}\frac{\nu_0\Delta}{v^2}\sqrt{\frac{\epsilon_0\Gamma_1^2}{\Delta^3}}.
\end{align}
In Fig.~3 of the main text we indicate $G^{*}_N$ for the positive and negative main peaks by arrows using the parameters given in Sec.~\ref{sec:theo_fit}.

At the crossover between regimes (ii) and (iii) the effective electron and hole tunneling rates are equal and the Andreev contribution to the conductance becomes resonant and reaches a maximum of order $2e^2/h$.
In regime (iii) both contributions to the differential conductance decrease with tunneling strength. This peculiar feature arises because of the strong energy dependence of the density of states in the superconducting tip and this regime has a sizable extension only when $ u^2$ and $v^2$ differ by several orders of magnitude. 
In the experiment this regime is presumably limited to very strong tunneling, $G_N>0.1(2e^2/h)$, where our approach ceases to be valid as the peak width becomes of the order of $\epsilon_0$. Furthermore the extension of this regime is too narrow to observe a decreasing peak height.

The thermal peak $\beta_-$ originates entirely from single-particle tunneling and its peak height simply follows the single-particle contribution to $\alpha_+$ up to a prefactor $\Gamma_2/\Gamma_1$ in regimes (i) and (ii). The remaining peaks $\alpha_-$ and $\beta_+$ have the same qualitative behavior as $\alpha_+$ and $\beta_-$ although with different regime boundaries, which are obtained from Eqs.~(\ref{Gn1}) and (\ref{Gn2}) by interchanging $u\leftrightarrow v$. The peak heights of all four peaks are shown in Fig.~\ref{fig:all_peaks_theo}(a) as a function of normal state conductance.

\subsection{Lineshape of the Shiba resonance}\label{sec:app_lineshape}

\begin{figure}[tb]
 \begin{center}
\includegraphics[width=.45\textwidth]{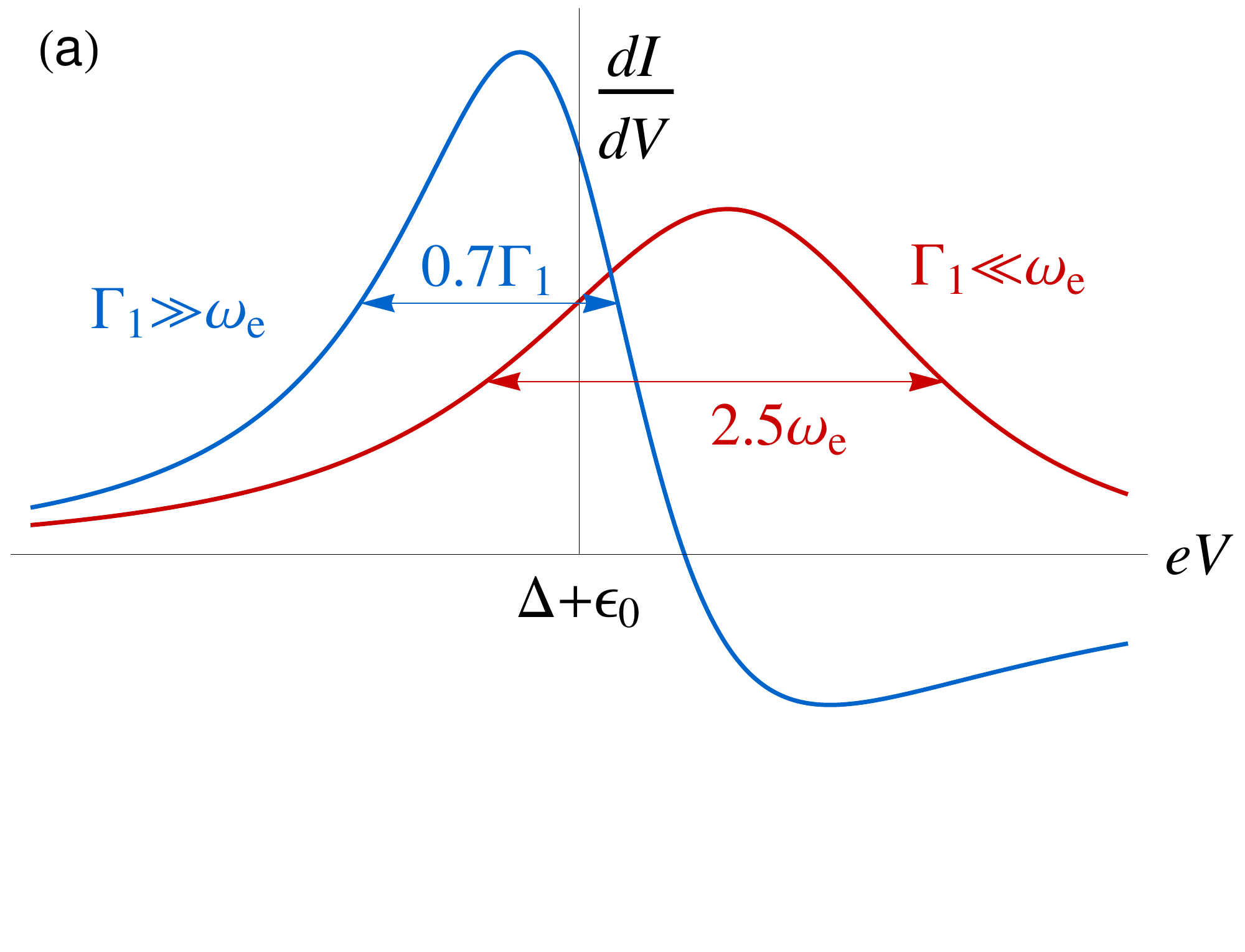} 
\includegraphics[width=.48\textwidth]{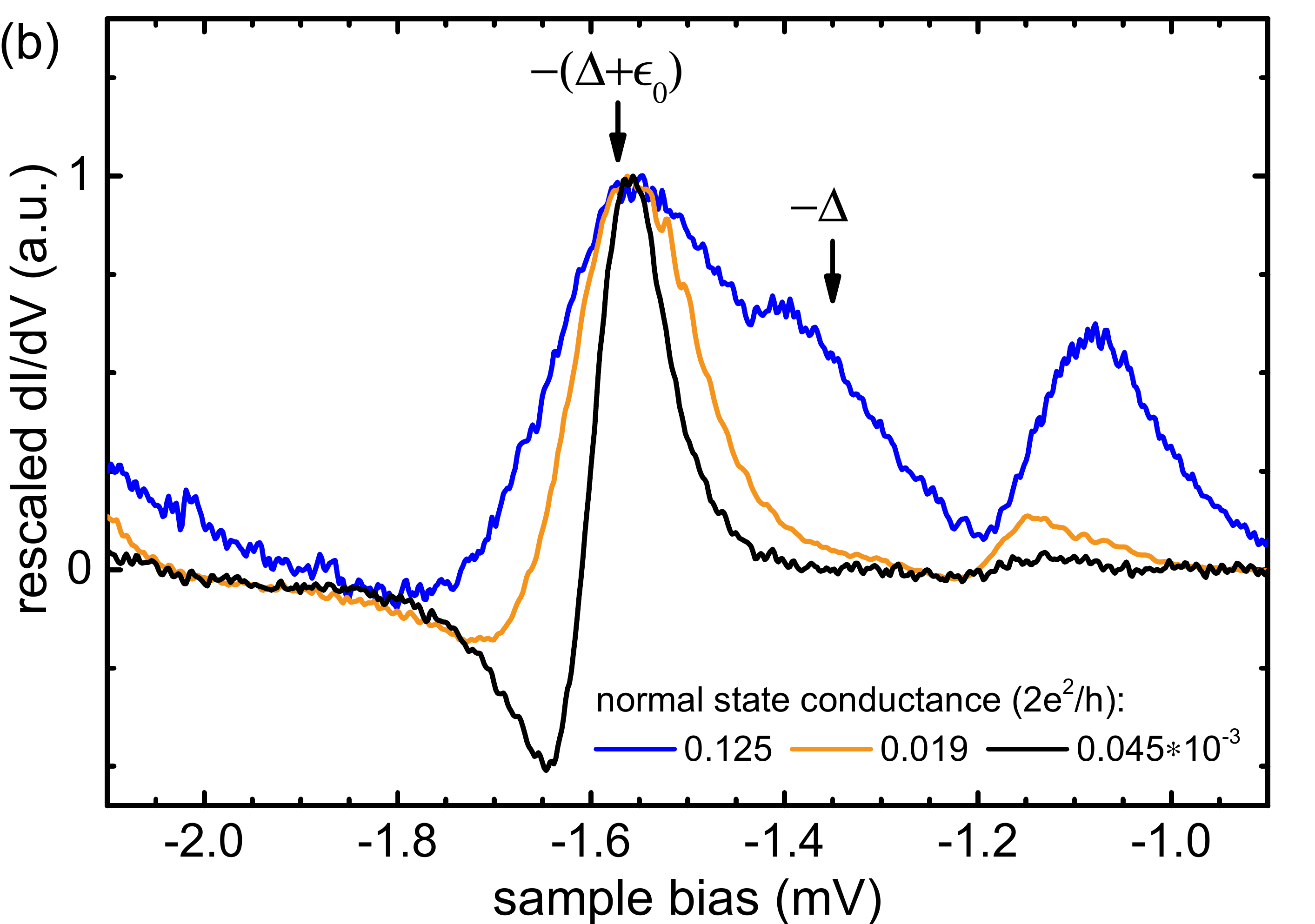}
\end{center}
\caption{(a) Lineshape of the differential conductance given by Eq.~(\ref{conductance_integral}) as a function of bias voltage near the main Shiba resonance at positive bias. The lineshapes differ qualitatively between the linear ($\Gamma_1\gg\omega_e$, blue) and sublinear regimes ($\Gamma_1\ll\omega_{e}$, red). Note that single-particle and Andreev currents give rise to identical lineshapes within the approximations of our analysis.
(b) Experimental $dI/dV$ traces of the Shiba peak near $eV=-(\Delta+\epsilon_0)\simeq -1.57\,\mathrm{meV}$ in the linear (black) and sublinear regime (blue). The spectra are normalized to the peak maximum. In the single-electron regime a clear negative differential resistance is visible around $-1.65\,\mathrm{mV}$.
}\label{fig:lineshape}
\end{figure}

In this section we analyze how the lineshape of the Shiba conductance peak is affected by the tunneling strength with important implications for fitting experimental $dI/dV$ traces. In the linear regime the single-particle current in Eq.~(\ref{app_current_relaxation}) can be approximated near $eV=\Delta+\epsilon_0$ by a convolution of the BCS density of states and a Lorentzian of width $\Gamma_1$
\begin{align}
 I_{\rm lin}(V)=2\pi eu^2t^2 \int \frac{d\omega}{2\pi\hbar} \rho(\omega-eV)\frac{\Gamma_1}{(\omega-\epsilon_0)^2+\Gamma_1^2/4}.
\end{align}
This expression has been used previously \cite{ji08,franke11} to fit experimentally measured Shiba resonances. The intrinsic Lorentzian lineshape can be obtained by numerical deconvolution of the data with the BCS density of states of the tip.

In the sublinear regime where the broadening of the Shiba resonance is determined by the tunnel coupling $\Gamma_e$ to the tip the current reads
\begin{align}
 I_{\rm sublin}(V)= 2\pi eu^2t^2 \int \frac{d\omega}{2\pi\hbar} \rho(\omega-eV)\frac{\Gamma_1}{(\omega-\epsilon_0)^2+[\pi u^2t^2\rho(\omega-eV)]^2}.
\end{align}
The tip density of states now also enters the width of the resonance and this expression does not have the form of a convolution. Consequently the lineshape of the Shiba resonance changes qualitatively from linear to sublinear regime as shown in Fig.~\ref{fig:lineshape}(a), where we plot the voltage dependence of the differential conductance given by Eq.~(\ref{conductance_integral}).
Strikingly, the maximum can occur above or below the threshold $eV=\Delta+\epsilon_0$ depending on the tunneling strength. 
This shift must be accounted for when determining the Shiba state energy by fitting experimental lineshapes. In addition, a characteristic negative differential conductance dip occurs for $\Gamma_1\gg\omega_{e}$ but is absent in the opposite regime. The disappearance of this dip in the measured lineshape provides a further indication of the crossover between weak and strong tunneling regimes. In the experiment the negative differential conductance dip indeed vanishes as tunneling strength is increased from linear to sublinear regime as shown in Fig.~\ref{fig:lineshape}(b). The peak position does not shift which we attribute to a slight suppression of the tip gap with increasing current.

Our analytical results in the previous subsections refer to the differential conductance exactly at the threshold voltages $e|V|=\Delta\pm\epsilon_0$. The shift of the maximum away from the threshold yields a somewhat larger peak height. Given that the peak heights vary by orders of magnitude in the experiment this deviation of at most 35\% only affects details but is inessential to our central results. Note that we calculate the actual peak height and not the threshold values in the quantitative comparison to the experimental data in Sec.~\ref{sec:theo_fit}.

\section{Theoretical analysis of the experimental data}

Our analysis implies that the subgap transport provides insight into the population dynamics of the Shiba state, as governed by the competition between tunneling and quasiparticle relaxation. We now fit the data against the results of our model and extract the quasiparticle lifetime in the Shiba state due to thermal relaxation processes. Besides demonstrating the validity of our description, this also yields valuable information about the dominant transport mechanisms in experiment as a function of the tunnel coupling between tip and sample.

\begin{figure}[t]
 \begin{center}
\includegraphics[width=.45\textwidth]{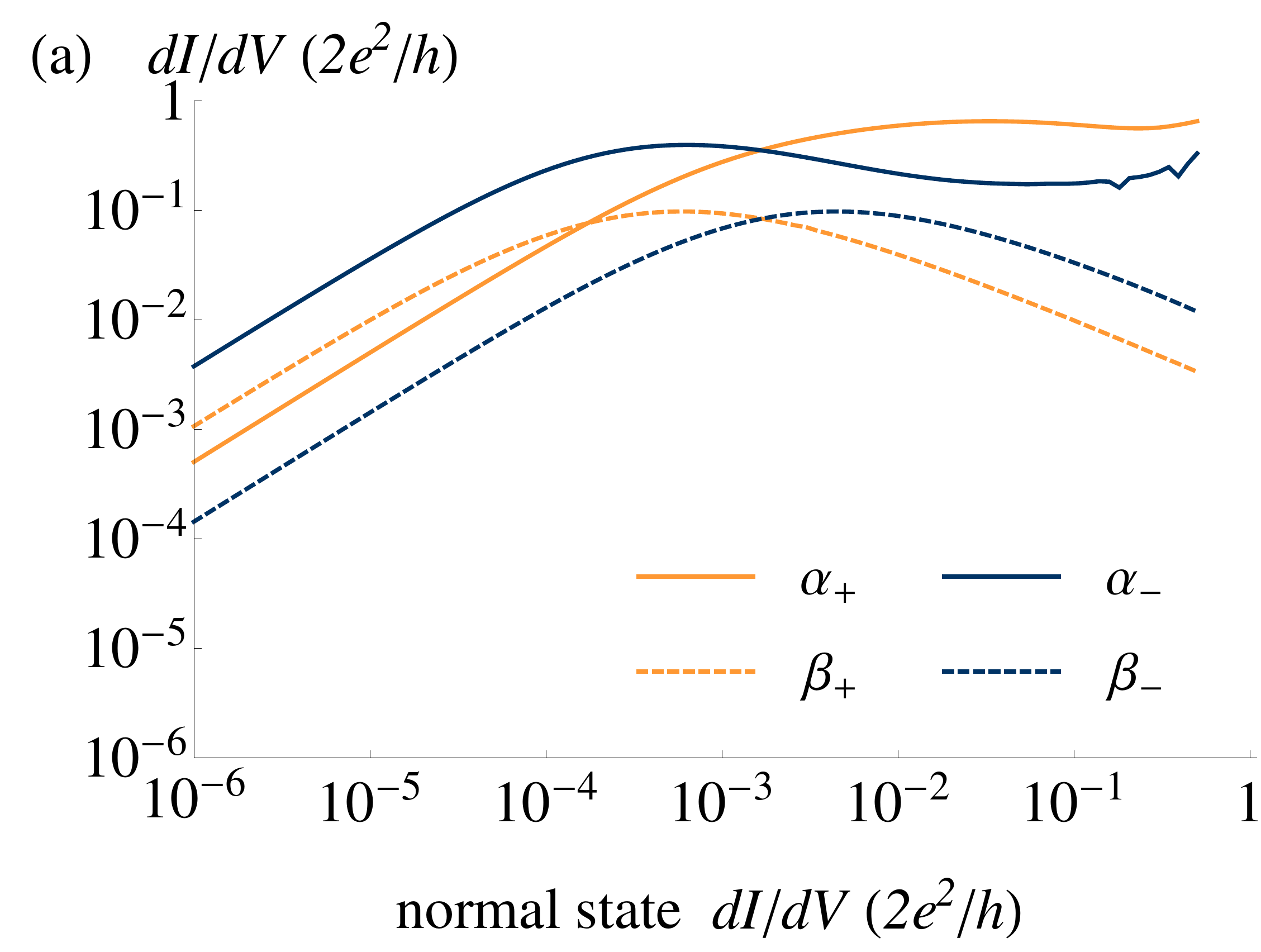} 
\includegraphics[width=.48\textwidth]{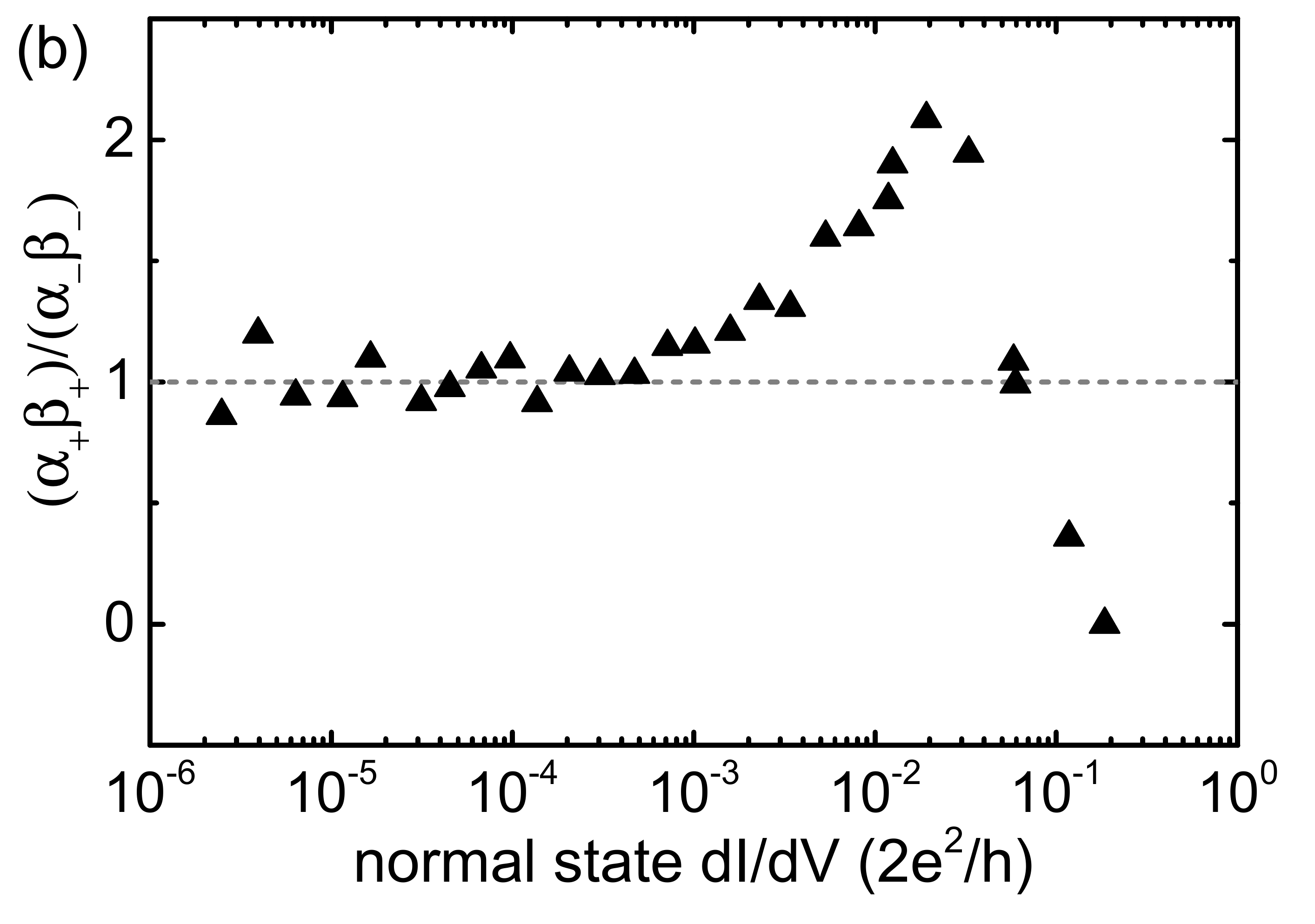}
\end{center}
\caption{(a) Differential conductance peaks at the thresholds {\em vs.} normal state conductance according to Eqs.~(\ref{IR_app}) and (\ref{IA_app}). We have used the parameters mentioned in Sec.~\ref{sec:theo_fit}.
(b)
Plot of $(\alpha_+\beta_+)/(\alpha_-\beta_-)$ \textit{vs.} conductance as extracted from Fig.~3 in the main manuscript. The ratio stays constant up to $10^{-3}\,\mathrm{G}_0$, where the sub-linear regime sets in.
}\label{fig:all_peaks_theo}
\end{figure}

\subsection{System parameters and theory fit}\label{sec:theo_fit}

\begin{figure}[t]
 \begin{center}
\includegraphics[width=.45\textwidth]{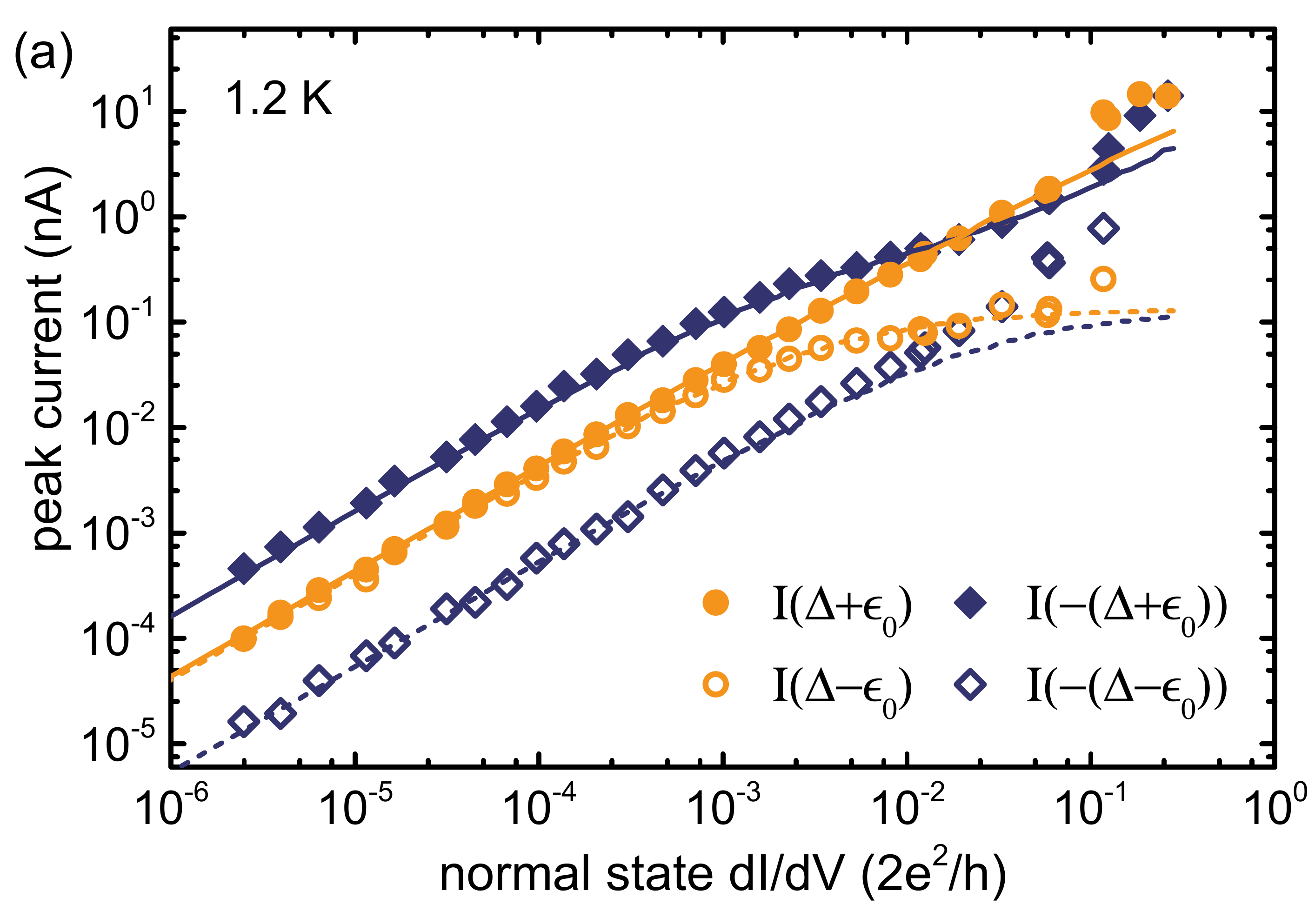} 
\includegraphics[width=.45\textwidth]{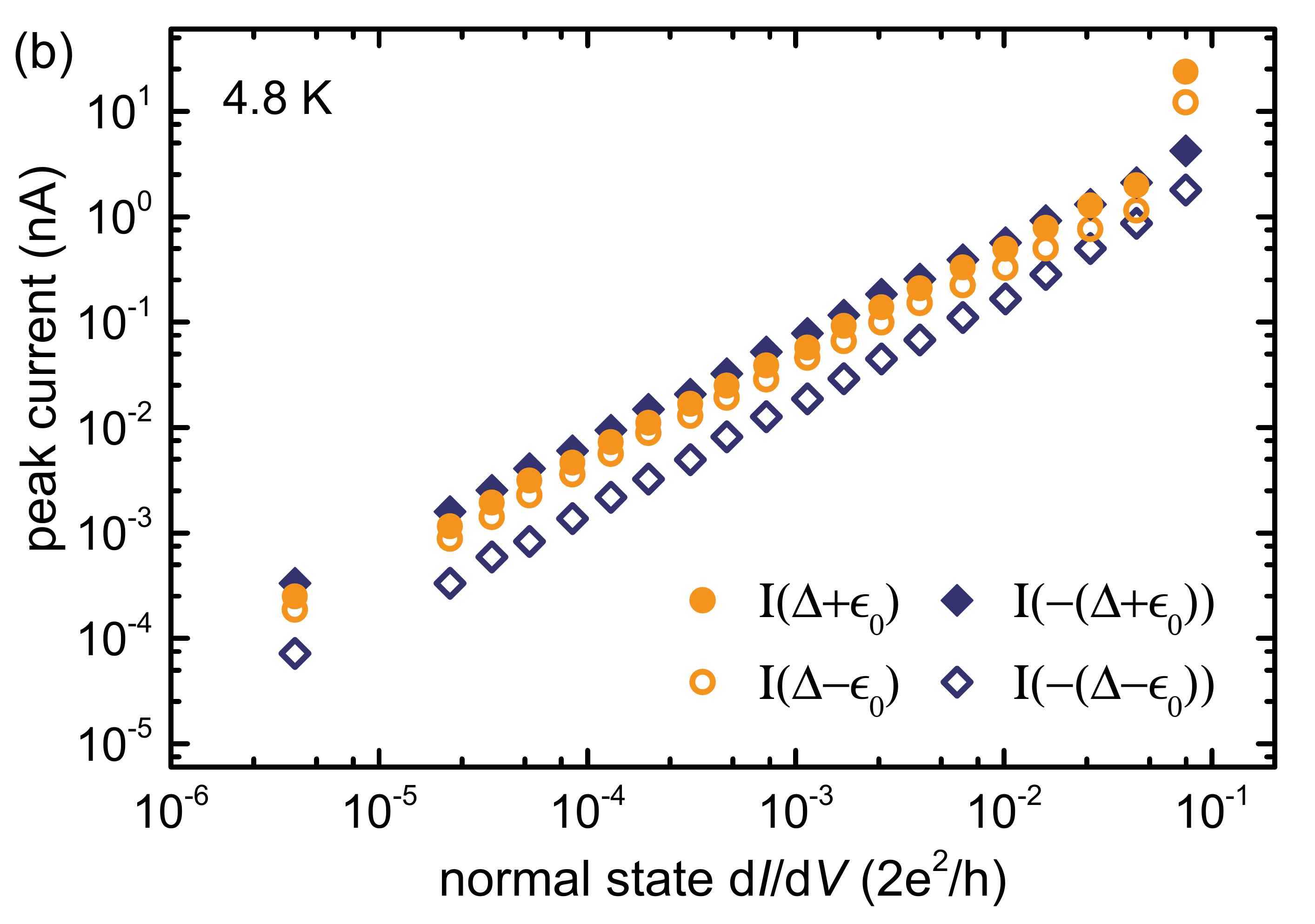} 
\end{center}
\caption{(a) Measured current at $1.2\,\mathrm{K}$ at the two main Shiba peaks ($eV=\pm(\Delta+\epsilon_0)$), and the corresponding two thermal peaks ($eV=\pm(\Delta-\epsilon_0)$). The lines show the theoretical curves using the parameters mentioned in the text. The parameters are the same as for the fit of the differential conductance in Fig.~3 of the main text.
(b) Measured current at $4.8\,\mathrm{K}$. 
}\label{fig:current}
\end{figure}

Here we provide details of the theoretical fits to the conductance and current at the Shiba peaks as a function of normal state conductance, as shown in Fig.~3 of the main text and Fig.~\ref{fig:current}(a). Several physical parameters can be extracted directly from the measured data without fitting. 
The Shiba energy can be determined from the location of the Shiba peaks as a function of bias voltage. For instance, the separation between the two positive bias peaks $\alpha_+$ and $\beta_+$ is $2\epsilon_0$. The same holds for $\alpha_-$ and $\beta_-$. From the data we estimate $\epsilon_0\simeq 0.22\,\mathrm{meV}$ for the lowest Shiba level. According to Eqs.~(\ref{alphaplus}) and (\ref{alphaminus}) valid in the linear regime we furthermore obtain $(u/v)^2=\alpha_+/\alpha_-\simeq 0.13$. Using Eq.~(\ref{betaplus}) in addition yields $\Gamma_1/\Gamma_2=\alpha_+/\beta_-\simeq 4$. Finally, we can accurately determine $\Gamma_2= 0.9(3)\,\mu\mathrm{eV}$ from the saturation of single-particle processes as detailed in Sec.~\ref{sec:lifetime}. With this the dimensionless Nambu spinor component $u^2/\nu_0\Delta$ that describes the spectral weight of the Shiba state at the impurity site remains the only unknown parameter in our model. Our results also predict the relation $\alpha_+\beta_+/\alpha_-\beta_-=1$ in the linear regime, which we can use as an additional check of the robustness of our theoretical description. According to the data shown in Fig.~\ref{fig:all_peaks_theo}(b) this relation is satisfied remarkably well over more than two decades of normal state conductance, throughout the linear regime.

The theoretical conductance peak heights obtained numerically from Eqs.~(\ref{IR_app}) and (\ref{IA_app}) are plotted in Fig.~\ref{fig:all_peaks_theo}(a) as a function of the normal state conductance
\begin{align}
 G_N=\frac{4\pi^2\nu_0^2 t^2}{1+\pi^2\nu_0^2 t^2}.
\end{align}
While several features of the theoretical curves qualitatively agree with the experimental data in Fig.~3 of the main text there are also notable deviations. Most prominently, theory predicts a peak in $\alpha_-$ at intermediate tunneling strength absent in the experiment. We attribute this deviation to extrinsic broadening introduced by the measurement setup, {\it e.g.}, due to radio frequency noise. Indeed, in the low coupling limit we find a peak width $w\simeq 70\,\upmu\mathrm{eV}$ [see black curve in \ref{fig:lineshape}(b)] exceeding $\Gamma_1$ by more than an order of magnitude. To account for the broadening we convolute the theoretical $dI/dV$ curves with a Gaussian of width $w$ and plot the resulting peak heights in Fig.~3 of the main text. We find remarkable agreement with the experimental data over the entire range of normal state conductance and determine $u^2/\nu_0\Delta\simeq 0.23$ from the fit. We associate deviations for $\beta_-$ at large normal state conductances $G_N\gtrsim 0.02 (2e^2/h)$ with a multiple Andreev reflection resonance involving the second Shiba state at $\epsilon_1\simeq 0.77\,\mathrm{meV}$. A resonance occurs at $eV=-(\Delta+\epsilon_1)/2\simeq -1.06\,\mathrm{meV}$ and therefore overlaps with the thermal Shiba peak at $eV=-(\Delta-\epsilon_0)\simeq - 1.13\, \mathrm{meV}$ [see blue curve in Fig.~\ref{fig:lineshape}(b)].

In Fig.~\ref{fig:current}(a), we plot the current measured at the position of the conductance peaks together with the theoretical curves. These fits use the {\em same} parameters as for the conductance fits, including the extrinsic broadening. We again find excellent agreement which corroborates that our model calculation correctly captures the essential tunneling processes.

\subsection{Quasiparticle lifetime and relaxation mechanism}\label{sec:lifetime}

The quasiparticle (quasihole) lifetimes of the Shiba state are related to the inverse relaxation rates $\tau_{1/2}=\hbar/\Gamma_{1/2}$. These rates could in principle be determined from the linewidth of the Shiba resonance at weak coupling (see Sec.~\ref{sec:app_lineshape}). The measured linewidth, however, may be increased by an additional broadening from the measurement setup that sets the experimental energy resolution. The intrinsic linewidth of the data taken $T=1.2\,{\rm K}$ is most likely well below the resolution of the experiment (see also Sec.~\ref{sec:theo_fit}).

A more robust way to determine the lifetime focuses on the strong-tunneling regime where the data remains unaffected by the energy resolution. As discussed in Sec.~\ref{sec:app_discussion}, the single-particle current saturates when $\Gamma_1<\omega_e$ assuming a value of $I^s=e\Gamma_{1/2}/3\hbar$ at the main (thermal) thresholds.
The current measured at the thermal threshold $eV=\Delta-\epsilon_0$ shown in Fig.~\ref{fig:current}(b) indeed exhibits a plateau at strong tunneling. At $eV=\Delta-\epsilon_0$ we extract a saturation current of $ 0.07(2)\,\mathrm{nA}$ which yields relaxation rates $\Gamma_2= 0.9(3)\,\upmu\mathrm{eV}$ and $\Gamma_1\simeq 4\Gamma_2= 4(1)\,\upmu\mathrm{eV}$. The  corresponding lifetimes $\tau_1\simeq0.2\,\mathrm{ns}$ and $\tau_2\simeq0.7\,\mathrm{ns}$ are quoted in the main text. The current $I_{-(\Delta-\epsilon_0)}$ at the other thermal threshold does not saturate because of additional subgap features at strong tunneling discussed in Sec.~\ref{sec:theo_fit}. 

Alternatively, we can determine $\Gamma_1$ from the current at the main threshold. Because of the additional Andreev contribution, the current $I_{-(\Delta+\epsilon_0)}$ exhibits a shoulder instead of a plateau when the single-particle current saturates [cf.\ Fig.~\ref{fig:double_peak_theo}(b)]. The current at the shoulder is $0.4(2)\,\mathrm{nA}$ and thus $\Gamma_1= 5(3)\,\upmu\mathrm{eV}$, in agreement with the above value. While determining $\Gamma_1$ from the shoulder at the main thresholds typically has a larger uncertainty, it is furthermore subject to a systematic error due to extrinsic broadening introduced in the measurement as detailed in Sec.~\ref{sec:theo_fit}. Note that the shoulder is absent in $I_{\Delta+\epsilon_0}$ because the Andreev current dominates already when the single-particle current saturates. This behavior is well captured by a quantitative comparison between theory and experiment in Fig.~\ref{fig:current}(a).

In order to gain insight into the relevant relaxation mechanisms in our system, we also study transport at a higher temperature of $T=4.8\,\mathrm{K}$. While the qualitative temperature dependence of the relaxation rates $\Gamma_{1/2}$ indicates thermal relaxation as discussed in the main text, a more quantitative analysis is required to assess the relevance of particular relaxation processes. At the higher temperature relaxation is strong enough that the single-particle current does not saturate in the tunneling regime $G_N\ll 2e^2/h$ as shown by the data in Fig.~\ref{fig:current}(b). We can instead determine $\Gamma_1+\Gamma_2$ from the peak width in the weak-coupling regime which we find to be $\simeq 0.16\,\mathrm{meV}$. Assuming the extrinsic broadening is the same for both temperatures, we estimate $\Gamma_1+\Gamma_2\simeq 0.2\,\mathrm{meV}$ [note that the unbroadened peak width is $0.7(\Gamma_1+\Gamma_2)$]. From the conductance peaks shown in the inset of Fig.~3 of the main text, we find $\alpha_-/\beta_+=\Gamma_1/\Gamma_2\simeq 1.6$ in the linear regime and thus $\Gamma_1\simeq 120\,\upmu\mathrm{eV}$ which correspond to $\tau_1\simeq 6\, \mathrm{ps}$. Hence we find the ratio of relaxation rates at the two temperatures $\Gamma_1(4.8\,\mathrm{K})/\Gamma_1(1.2\,\mathrm{K})\simeq 35$. 

In Sec.~\ref{sec:relaxation} we have discussed quasiparticle transitions between Shiba state and quasiparticle continuum as possible relaxation mechanisms. 
Based on the corresponding relaxation rate in Eq.~(\ref{gamma1_temp}), however, we would expect a ratio $\Gamma_1(4.8\,\mathrm{K})/\Gamma_1(1.2\,\mathrm{K})\sim 10^4$ (using $\epsilon_0=0.20\,{\rm meV}$ and $\Delta=1.21\,{\rm meV}$ at $4.8\,\mathrm{K}$), reflecting the exponential suppression of thermal relaxation at low temperatures. In view of the large discrepancy to the experimental value we can exclude direct transitions to the quasiparticle continuum as the dominant relaxation process. An alternative scenario involves a cascade of transitions first to the next Shiba state at $\epsilon_1=0.77\,{\rm meV}$ ($\epsilon_1=0.68\,{\rm meV}$ at $4.8\,\mathrm{K}$) and subsequent relaxation to the third Shiba state and the continuum. A rough estimate using Eq.~(\ref{gamma1_temp}) with $\epsilon_1$ instead of $\Delta$ yields a ratio $\Gamma_1(4.8\,\mathrm{K})/\Gamma_1(1.2\,\mathrm{K})\sim 10^2$. This result is much closer to the experimental value though a more detailed analysis of the population dynamics of the various Shiba states is required for a quantitative comparison. Note that a slightly higher temperature than $1.2\,\mathrm{K}$ could account for a significant part of this deviation.

%
%

%
%
%
%
%
%
%
%
%
%

\end{document}